\documentclass[%
 aip,
 apl,
 amsmath,amssymb,
 reprint,%
]{revtex4-1}
\usepackage{amsmath}    % need for subequations
\usepackage{amsfonts}
\usepackage{bm}
\usepackage{amssymb}
\usepackage{graphicx}   % need for figures
\usepackage[usenames,dvipsnames]{xcolor}      % use if color is used in text
\usepackage{comment}
\usepackage{siunitx}
\usepackage{enumitem}
\usepackage{upgreek}
\usepackage{color}

\usepackage[utf8]{inputenc}
\usepackage[T1]{fontenc}
\usepackage{mathptmx}
\usepackage{etoolbox}

\newcommand{\trans}{\mathsf{T}}
\renewcommand{\vec}[1]{\mathbf{#1}}

\newcommand{\spec}{\mbox{\rm spec}}
\renewcommand{\eqref}[1]{Eq.\ (\ref{#1})}

\newcommand*{\citen}[1]{%
  \begingroup
    \romannumeral-`\x % remove space at the beginning of \setcitestyle
    \setcitestyle{numbers}%
    [\cite{#1}]%
  \endgroup   
}

\makeatletter
\def\@email#1#2{%
 \endgroup
 \patchcmd{\titleblock@produce}
  {\frontmatter@RRAPformat}
  {\frontmatter@RRAPformat{\produce@RRAP{*#1\href{mailto:#2}{#2}}}\frontmatter@RRAPformat}
  {}{}
}%
\makeatother
\begin{document}

\title{Tutorial: Classifying Photonic Topology Using the Spectral Localizer and Numerical $K$-Theory} %Title of paper

\author{Alexander Cerjan}
\email[]{awcerja@sandia.gov}
\affiliation{Center for Integrated Nanotechnologies, Sandia National Laboratories, Albuquerque, New Mexico 87185, USA}

\author{Terry A.\ Loring}
%\email[]{loring@math.unm.edu}
\affiliation{Department of Mathematics and Statistics, University of New Mexico, Albuquerque, New Mexico 87131, USA}

\date{\today}

\begin{abstract}
Recently, the spectral localizer framework has emerged as an efficient approach to classifying topology in photonic systems featuring local nonlinearities and radiative environments. In nonlinear systems, this framework provides rigorous definitions for concepts like topological solitons and topological dynamics, where a system's occupation induces a local change in its topology due to the nonlinearity. For systems embedded in radiative environments that do not possess a shared bulk spectral gap, this framework enables the identification of local topology and shows that local topological protection is preserved despite the lack of a common gap. However, as the spectral localizer framework is rooted in the mathematics of $C^*$-algebras, and not vector bundles, understanding and using this framework requires developing intuition for a somewhat different set of underlying concepts than those that appear in traditional approaches to classifying material topology. In this tutorial, we introduce the spectral localizer framework from a ground-up perspective, and provide physically motivated arguments for understanding its local topological markers and associated local measure of topological protection. In doing so, we provide numerous examples of the framework's application to a variety of topological classes, including crystalline and higher-order topology. We then show how Maxwell's equations can be reformulated to be compatible with the spectral localizer framework, including the possibility of radiative boundary conditions. To aid in this introduction, we also provide a physics-oriented introduction to multi-operator pseudospectral methods and numerical $K$-theory, two mathematical concepts that form the foundation for the spectral localizer framework. Finally, we provide some mathematically oriented comments on the $C^*$-algebraic origins of this framework, including a discussion of real $C^*$-algebras and graded $C^*$-algebras that are necessary for incorporating physical symmetries. Looking forward, we hope that this tutorial will serve as an approachable starting point for learning the foundations of the spectral localizer framework.
\end{abstract}
% final intro paragraph

\pacs{}% insert suggested PACS numbers in braces on next line

\maketitle %\maketitle must follow title, authors, abstract and \pacs

\section{Introduction}

Since the seminal works of Haldane and Raghu \cite{haldane_possible_2008,raghu_analogs_2008} demonstrated that topological phenomena can manifest in any material system governed by a wave equation, the ideas of topological physics have excited the photonics community \cite{wang_observation_2009} both to use photonic platforms to explore new fundamental concepts as well as for leveraging topology's benefits in photonic devices. In particular, topological photonic systems are both guaranteed to exhibit localized states at their boundaries or corners and these states' existence are robust against fabrication imperfections, yielding an enticing suite of properties for enhancing light-matter interactions \cite{ota_active_2020} and routing quantum information \cite{lodahl_chiral_2017}. Early, and still ongoing, efforts in the field of topological photonics have focused on finding systems that realize non-reciprocal topological phenomena at technologically relevant wavelengths \cite{oka_photovoltaic_2009,kitagawa_topological_2010,hafezi_robust_2011,umucalilar_artificial_2011,lindner_floquet_2011,kitagawa_observation_2012,fang_realizing_2012,kraus_topological_2012,rechtsman_photonic_2013,khanikaev_photonic_2013,hafezi_imaging_2013,klembt_exciton-polariton_2018,dutt_single_2020} despite the present lack of materials with a strong magneto-optical response at those same wavelengths. In addition, the community has also explored other classes of topology that do not require breaking time-reversal symmetry \cite{lu_weyl_2013,wu_scheme_2015,lu_experimental_2015,blanco-redondo_topological_2016,barik_topological_2018,mittal_topological_2018,noh_topological_2018,peterson2018,blanco-redondo_topological_2018,chen_direct_2019,xie_visualization_2019,mittal_photonic_2019,ota_photonic_2019,smirnova_third-harmonic_2019,barik_chiral_2020,parappurath_direct_2020,cerjan_observation_2020,kim_multipolar_2020,vaidya_observation_2020,arora_direct_2021,kruk_nonlinear_2021,dai_topologically_2022,hauff_chiral_2022,jorg_observation_2022}, but whose protected states are either reciprocal or zero-dimensional cavity states. More recently, the field has begun to shift its focus towards using topology in photonic devices, creating topological lasers \cite{bahari_nonreciprocal_2017,st-jean_lasing_2017,bandres_topological_2018,zeng_electrically_2020,yang_spin-momentum-locked_2020,shao_high-performance_2020,bahari_photonic_2021,dikopoltsev_topological_2021,yang_topological-cavity_2022} as well as systems for controlling quantum light \cite{rechtsman_topological_2016,barik_topological_2018,mittal_topological_2018,barik_chiral_2020,parappurath_direct_2020,arora_direct_2021,dai_topologically_2022,hauff_chiral_2022}.

Traditionally, topological phenomena are identified via invariants calculated using the band structure and Bloch eigenstates of a gapped (i.e., insulating) crystalline system, and these invariants cannot change without first closing the bulk band gap \cite{hasan_colloquium_2010,asboth_short_2016}. As such, a system's invariants are protected against disorder that is not strong enough to close the band gap. Moreover, bulk-boundary correspondence \cite{Bellissard1994BulkBoundary,Elgart2005Bulk_boundary} guarantees that a system with a non-trivial invariant exhibits some set of boundary- or corner-localized states. Thus, topological band theory is ideally suited to predicting the edge transport properties of large systems without needing to ensure any particular edge termination and regardless of strong disorder along the boundary or interface. In other words, such an approach is ideal for predicting the behavior of condensed matter systems.

However, the frontiers of topological photonics are diverging from those of condensed matter physics due to the different physical phenomena available in, as well as the different desired applications of, each platform. In particular, photonic systems can exhibit local mean-field nonlinearities \cite{smirnova_nonlinear_2020}, arbitrarily tailorable geometry that enables small system sizes \cite{schulz_topological_2021,zhang_printing_2024}, and non-Hermiticity both through radiative losses as well as gain and absorption \cite{poo_experimental_2011,shen_topological_2018,kawabata_symmetry_2019,bergholtz_exceptional_2021}. Moreover, these phenomena are central to many photonic devices seeking to leverage topological phenomena. For example, it can be advantageous to reduce a device's on-chip footprint, which necessitates understanding finite size effects, or make surface emitting light sources, which requires the inclusion of radiative losses to a gapless medium. Unfortunately, photonic systems exhibiting these phenomena are ill-suited to be classified using topological band theory, either because the effects are local, or the system is small, such that a band theoretic description is not applicable; or because the full photonic system, including the radiative environment, lacks a spectral gap so band theory would not predict any global topological protection. 

In the last few years, a topological classification framework based on the system's \textit{spectral localizer} has emerged \cite{loringPseuspectra,LoringSchuBa_odd,LoringSchuBa_even}, which is able to identify photonic and material topology in realistic systems \cite{cerjan_operator_Maxwell_2022,cerjan_local_2024} regardless of whether a system has radiative boundaries \cite{dixon_classifying_2023,wong_classifying_2024} or local non-linearities \cite{wong_probing_2023}. This framework provides a position-space picture of topology, yielding spatially resolved local topological markers and a local measure of topological protection. As such, the spectral localizer framework offers the possibility of aiding in the advancement of the field of topological photonics as it explores nonlinearities and finite system-size effects. However, the spectral localizer and its associated local markers are rooted in the mathematics of $C^*$-algebras and cannot be expressed in the theories of geometry and vector bundles that are traditionally used to describe material topology. As such, the relevant formulae for using the spectral localizer framework initially appear quite foreign and do not suggest an immediate physical interpretation despite their utility. Moreover, the development of the spectral localizer framework is spread out over the physics and mathematics literature, yielding a daunting task for anyone interested in advancing its underlying theory or making use of the approach.

In this tutorial, we provide a physically motivated description of the spectral localizer framework and show examples of its use across a variety of different classes of topology, including those from the Altland-Zirnbauer classification \cite{schnyder2008,kitaev2009,ryu2010topological} as well as those of crystalline origin. In doing so, we both show how this framework provides quantitative predictions of a system's topological robustness, as well as demonstrate how bulk-boundary correspondence manifests. As part of this tutorial, we also introduce two main mathematical topics key to the spectral localizer: multi-operator pseudospectral methods and numerical $K$-theory. Multi-operator pseudospectral methods are an approach to finding approximate joint eigenvectors for non-commuting operators, and these methods form the basis for understanding bulk-boundary correspondence in the spectral localizer framework, though they have utility beyond the study of topological materials. Numerical $K$-theory is the study of deriving formula for topological invariants that yield efficient computational implementations; the local markers calculated using the spectral localizer are an example of numerical $K$-theory. Looking forward, we hope that this tutorial provides an approachable on-ramp for members of both the physics and mathematics communities to use the spectral localizer framework to classify topological systems.

\subsection{Preliminaries and terminology}

Overall, this tutorial assumes that the reader has some familiarity with the study of topological physics, but makes no assumptions on the reader's mathematical background beyond a standard introduction to linear algebra and the usual facts about linear operators and Hilbert spaces used in modern physics. Proofs for many mathematical details are instead provided in the associated references. The only exception to this assumption is in the self-contained Sec.~\ref{sec:whyWorks}, where some of the $C^*$-algebraic underpinnings of the spectral localizer framework are presented in a form for a reader with a mathematically oriented background.

There are also two overarching terminology issues that arise in the study of topological photonics. First, this tutorial uses the phrase \textit{material topology}, or variants of this, to refer to the possible topological behavior of any natural or artificial material, regardless of whether it is comprised of atoms, molecules, artificial atoms, or any other form of decoration that serves as a change in a system's spatial potential. Second, this tutorial will use the condensed matter physics terminology of a system's \textit{occupied} states or bands to refer to those states or bands of a system whose eigen-energies are less than some chosen energy of interest. Of course, in natural materials whose topology arises in their many-body electron wave function, the energy of interest is the Fermi energy and all of the (single particle) states with energies less than this are necessarily occupied at zero temperature. In contrast, in photonic systems, a structure's topological phenomena can be observed by exciting the system at a single frequency, without carefully populating the system at other frequencies. Indeed, as photons do not directly interact, their system Hamiltonians are linear (for linear materials) and do not change based on the system's occupation. Thus, a photonic system may exhibit several disparate spectral ranges with different non-trivial topology, any of which may be experimentally accessible. Nevertheless, as many formulae for topological invariants at a given energy $E$ are defined in terms of a system's states $|\psi_m \rangle$ with $E_m \le E$, it is useful to retain terminology that refers to this set of lower-energy states.

\subsection{Structure of the tutorial}

In Sec.\ \ref{sec:review}, we provide a brief review of topological band theory and related position-space topological invariants. In Sec.\ \ref{sec:SLintro}, we first provide an intuitive picture for building up the spectral localizer framework, including its local markers and measure of protection. This section then provides a set of general definitions for the framework and provides a number of examples across a variety of different classes of topology and system dimensions. In Sec.~\ref{sec:4}, we provide full discussions of the hyper-parameter $\kappa$ that appears in the spectral localizer framework, as well as the manifestation of bulk-boundary correspondence, and efficient numerical implementations through numerical $K$-theory. In Sec.~\ref{sec:photon}, we show how the spectral localizer framework can be applied to photonic systems, and discuss some of the challenges associated with considering systems described by differential operators like Maxwell's equations. In Sec.~\ref{sec:whyWorks}, we provide some self-contained comments on the mathematical background of the spectral localizer framework. Finally, in Sec.~\ref{sec:outlook}, we provide some concluding remarks. Throughout the tutorial, we also directly acknowledge \underline{open questions} (marked as such) and other areas that require further study.

\section{Brief Review of Topological Classification Theories \label{sec:review}}

In this section, we provide a very brief review of standard methods for classifying material topology: topological band theory, global position-space invariants, and local topological markers. This review is intended as a reminder, not a detailed introduction (instead see Refs.~\citen{hasan_colloquium_2010,lu_topological_2014,chiu_classification_2016,asboth_short_2016,ozawa_topological_2019}), and is included so that similarities and differences of these methods with the spectral localizer framework can be highlighted in later sections. As such, this brief review focuses solely on how each framework identifies Chern materials (2D class A in the Altland-Zirnbauer classification table \cite{schnyder2008,kitaev2009,ryu2010topological}), as the distinctions between these approaches are similar across every topological class. Likewise, this section may also be safely skipped if a reader is already familiar with, or uninterested in, this background.

\subsection{Topological band theory}

Traditionally, material topology is classified using topological invariants defined in terms of quantities derived from a system's band structure, such as the Chern number \cite{thouless_quantized_1982}, winding number \cite{asboth_short_2016}, or multipole moment \cite{zak_berrys_1989,benalcazar2017quad}. Such band theory-based invariants are only defined for materials that possess a bulk band gap (i.e., insulators), and these invariants can only change when a sufficiently strong perturbation is added to the system that closes the bulk band gap. Thus, by definition, standard band theoretic topological invariants are global properties of a crystalline material, as they assume the material is infinite and periodic so that Bloch's theorem can be applied. Materials possessing non-trivial bulk topological invariants necessarily exhibit associated boundary-localized states, whose appearance is guaranteed by bulk-boundary correspondence \cite{Prodan_Schuba2016Bulk_Boundary_book,Bourne2017Bulk_Boundary_KK-theory,Kaufmann2024_index_defns}.

A 2D material's Chern number $C_{E}$ for an energy $E$ in a specified spectral gap, can be found from its Bloch eigenstates $|\psi_{m \vec{k}}\rangle = e^{i \vec{k} \cdot \vec{r}} |u_{m \vec{k}}\rangle$ as
\begin{equation}
    C_{E} = \frac{1}{2 \pi} \sum_{m=1}^{M_{\textrm{occ}}} \int_{\textrm{BZ}} i \nabla_{\vec{k}} \times \langle u_{m \vec{k}}| \nabla_{\vec{k}} | u_{m \vec{k}} \rangle d^2 \vec{k} \in \mathbb{Z}. \label{eq:bandC}
\end{equation}
Here, $|\psi_{m \vec{k}}\rangle$ is an eigenstate of the system's Bloch-periodic Hamiltonian at wavevector $\vec{k}$ and is a part of the $m$th band, $M_\textrm{occ}$ is the number of occupied bands below the spectral gap containing $E$, and the integral is taken over the entire first Brillouin zone. By directly specifying the energy in \eqref{eq:bandC}, we are emphasizing a crucial difference between topological materials as they appear in condensed matter physics and photonics: There is generally only a single relevant spectral gap in electronic insulators, i.e., at the Fermi energy, whereas a photonic material can generally be excited at any frequency, and may exhibit multiple bulk band gaps that can each possess a different topological invariant. Finally, as the topological protection predicted by band theory is given by the size of the bulk band gap surrounding $E$, topological band theory cannot be meaningfully applied to $E$ chosen in the spectral extent of the bulk bands.

\subsection{Global position-space invariants}

It is also possible to classify material topology using a system's position-space description, rather than its momentum-space description. 
%%%%%%%
Heuristically, such an approach must be possible, as the two descriptions are related by a Fourier transform, and Fourier transforms neither add nor remove information \cite{plancherel_contribution_1910}, they simply rearrange it. Thus, the aspects of a crystalline material that give rise to non-trivial topology, as determined by topological band theory, must also be detectable using invariants that leverage a system's position-space Hamiltonian $H$ rather than its Bloch periodic Hamiltonian $H(\vec{k})$. In general, there are two different approaches to classifying material topology in position space: global invariants and local markers. Global position-space invariants always make use of a system Hamiltonian with periodic boundary conditions (PBC), but this Hamiltonian represents a large volume of the material rather than a single unit cell with Bloch periodic boundary conditions as is used by topological band theory. Theories of local topological markers can be constructed using either systems with open boundary conditions (OBC) or PBC, and classify the system's topology at a chosen location.

The canonical example of a global position-space invariant is the Bott index \cite{hastings_topological_2011}, 
\begin{equation}
    \textrm{Bott}_{E} = \textrm{Re}\left[\frac{1}{2\pi i} \textrm{Tr} \left( \log \left(U_X U_Y U_X^\dagger U_Y^\dagger \right) \right) \right] \in \mathbb{Z} \label{eq:bottC}
\end{equation}
which identifies the same topological phenomena as the Chern number. Here, $E$ is again assumed to be in a spectral gap, and 
\begin{subequations}
\begin{align}
    U_X = & \Psi_{\textrm{occ}}^\dagger e^{\frac{2 \pi i}{l_x} X} \Psi_{\textrm{occ}}, \\
    U_Y = & \Psi_{\textrm{occ}}^\dagger e^{\frac{2 \pi i}{l_y} Y} \Psi_{\textrm{occ}},
\end{align}
\end{subequations}
are the periodic (Resta) position operators \cite{resta_quantum-mechanical_1998} written in the subspace of the system's occupied states with energies below the spectral gap at $E$, $X$ and $Y$ are the standard (non-periodic) single-particle position operators, and $l_x$ and $l_y$ are the lengths of the system in the in-plane directions. In other words, $\Psi_{\textrm{occ}} = [|\psi_1\rangle,...,|\psi_m\rangle,...,|\psi_M\rangle]$, with $H |\psi_m\rangle = E_m |\psi_m\rangle$ and $E_m < E$, i.e., $\Psi_{\textrm{occ}}$ is the rectangular matrix whose columns are the eigenstates of the large, periodic Hamiltonian with energies below the chosen spectral gap. Since the discovery of the Bott index, related Bott-like indexes have been subsequently developed for a broad range of different classes of topology \cite{huang_quantum_2018,lin_real-space_2021,benalcazar_chiral-symmetric_2022}. There are also other frameworks for global position-space invariants that were discovered at the about same time as the Bott index and take a very different approach \cite{Prodan2010EntanglementSpectrum}.

When calculating a periodic material's topology using \eqref{eq:bottC}, the size of $H$ that is typically required (before reaching the periodic boundary condition) to guarantee accurate classification is related to ensuring that the material's topological band inversion can be resolved. In other words, a finite $H$ with PBC is sampling wavevector space every $\delta k_j = 2\pi/l_j$, where $l_j$ is the length of the system in the $j$th direction, so $l_j$ must be large enough to ensure the region of wavevectors where the bands have inverted \cite{bansil_colloquium_2016} is sufficiently sampled by the Bott index. Thus, heuristically, the Bott index is trading repeated calculations of the Bloch Hamiltonian $H(\vec{k})$'s spectrum and eigenstates at different $\vec{k}$ that are required to perform the full integral in \eqref{eq:bandC} with a single calculation of the spectrum and eigenstates of the larger $H$. Additionally, the matrices $U_X$ and $U_Y$ are generally dense.

\subsection{Local topological markers}

In contrast to topological band theory and global position-space invariants, local topological markers are calculated at a specified position $\vec{x}$, as well as a specified energy. Thus, a non-uniform system's local markers can vary across the system, indicating regions with different material topology. As such, theories of local markers can be applied to aperiodic materials, such as quasicrystals \cite{tran_topological_2015,bandres_topological_2016,fulga_aperiodic_2016,grossi_e_fonseca_quasicrystalline_2023} and amorphous structures \cite{mitchell_amorphous_2018,mitchell_real-space_2021}, as well as disordered systems or heterostructures \cite{sahlberg_quantum_2023}, without alteration. The first local marker was derived by Kitaev in 2006 \cite{kitaev2006anyons}, and Bianco and Resta produced a seminal study on the topic in 2011 \cite{bianco_mapping_2011}. Both of these initial local markers identify Chern topology in finite systems with open boundaries, but a variety of local markers for different classes of topology have been subsequently derived \cite{kubota_controlled_2017,bourne_non-commutative_2018,li_local_2019,caio_topological_2019,sykes_local_2021,velury_topological_2021,dornellas_quantized_2022,hannukainen_local_2022,chen_universal_2023,munoz-segovia_structural_2023,herzog-arbeitman_hofstadter_2023,kim_replica_2023,mondragon-shem_robust_2024,bau_local_2024,doll_local_2024,herzog-arbeitman_interacting_2024,velury_global_2024}.

\begin{figure}[t]
    \centering
    \includegraphics{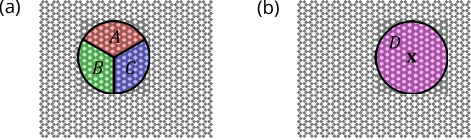}
    \caption{Schematic of the three regions used in the evaluation of the Kitaev local Chern marker \eqref{eq:kitaev} (a) and the disk used by the Bianco-Resta local Chern marker \eqref{eq:bianco} (b).}
    \label{fig:kitaev}
\end{figure}

Kitaev's local Chern marker originates from considering spectral flow in a unitary matrix \cite{kitaev2006anyons} and can be understood as determining the system's response to a point-like magnetic field at the specified position $\vec{x}$ \cite{mitchell_amorphous_2018}. To do so, the Kitaev marker partitions a system into three spatial regions, typically labelled $A$, $B$, and $C$ with positive orientation, that all touch at $\vec{x}$, see Fig.\ \ref{fig:kitaev}a. This local Chern marker is calculated through the flow of the system's projector across the three regions
\begin{multline}
    \nu_{(\vec{x},E)}(A,B,C) \\ 
    = -12 \pi i \sum_{\vec{x}_j \in A} \sum_{\vec{x}_k \in B} \sum_{\vec{x}_l \in C} \left(P_{jk} P_{kl} P_{lj} - P_{jl} P_{lk} P_{kj} \right), \label{eq:kitaev}
\end{multline}
in which the system's protector onto its states with $E_m < E$ is
\begin{equation}
    P = \sum_{E_m < E} |\psi_m \rangle \langle \psi_m |,
\end{equation}
whose elements for a finite system are then
\begin{equation}
    P_{jk} = \sum_{E_m < E} \langle \vec{x}_j |\psi_m \rangle \langle \psi_m | \vec{x}_k \rangle,
\end{equation}
where $|\psi_m \rangle$ are eigenstates of the finite Hamiltonian with open boundaries and $| \vec{x}_j \rangle$ are position eigenstates localized entirely on site $j$. %, and whose $j$th element is $\langle \vec{x}_j |\psi_m \rangle$.
Note that the sign of \eqref{eq:kitaev} is flipped relative to the original definition as we are reviewing the result that uses the standard projector to the system's states below $E$ \cite{mitchell_real-space_2021}. So long as the three regions are sufficiently large, but do not extend to the material's boundary, $\nu_{(\vec{x},E)}(A,B,C)$ is approximately an integer and for crystalline systems with $\vec{x}$ chosen within the material bulk is approximately equal to \eqref{eq:bandC}. If the regions are instead chosen to include the material's boundary, $\nu_{(\vec{x},E)} = 0$ \cite{kitaev2006anyons}. Additionally, \eqref{eq:kitaev} is approximately invariant under changing the boundaries of the three regions for the same $\vec{x}$, so long as the regions remain sufficiently large and positively oriented.

Bianco and Resta's local Chern marker for 2D systems can be understood as the Fourier transform of \eqref{eq:bandC} and subsequently localized to a large-but-finite region of a material \cite{bianco_mapping_2011}. This marker can be expressed in terms of the system's projectors to both those states below and above the choice of $E$
\begin{equation}
    Q = \mathbf{1} - P
\end{equation}
as \cite{tran_topological_2015,marrazzo_locality_2017,grossi_e_fonseca_quasicrystalline_2023}
\begin{equation}
    \mathfrak{C}_{(\vec{x},E)}(D) = -\frac{4\pi}{A_D} \sum_{\vec{x}_i \in D} \textrm{Im} \left[ \sum_{j} \langle \vec{x}_i | x_Q | \vec{x}_j \rangle \langle \vec{x}_j | y_P | \vec{x}_i \rangle \right], \label{eq:bianco}
\end{equation}
in which
\begin{subequations}
\begin{align}
    \langle \vec{x}_i | x_Q | \vec{x}_j \rangle = & \sum_{k} Q_{ik} x_k P_{kj}, \\ 
    \langle \vec{x}_j | y_P | \vec{x}_i \rangle = & \sum_{k} P_{jk} y_k Q_{ki}.
\end{align}    
\end{subequations}
Here, $\mathbf{1}$ is the identity, and the local Chern marker is determined by integrating over a disk $D$ with area $A_D$ and center $\vec{x}$, with $\vec{x} = (x,y)$, see Fig.\ \ref{fig:kitaev}b. Similar to \eqref{eq:kitaev}, for crystalline materials $\mathfrak{C}_{(\vec{x},E)}(D)$ converges to $C_{E}$ from \eqref{eq:bandC} as the radius of $D$ increases, subject again to the same caveat that the disk does not contain any portion of the material's boundary. If $D$ instead includes the full finite system with open boundaries, $\mathfrak{C}_{(\vec{x},E)}(D) = 0$ \cite{bianco_mapping_2011}.

Despite their different formulations, the Kitaev marker and the Bianco-Resta marker both exhibit similar properties. Unlike band theoretic invariants or global position-space invariants, neither are guaranteed to be integers for finite choices of integration areas (i.e., $A$, $B$, and $C$ for \eqref{eq:kitaev} or $D$ for \eqref{eq:bianco}), though both converge to the material's Chern number as the integration area is increased. Moreover, both local markers, as well as the Bott index, typically involve calculating all of a system's eigenstates with energies below $E$, which can be numerically expensive given that these formulations require using Hamiltonians that represent large volumes of material. In some cases, it is possible to circumvent finding a system's states through the kernel polynomial method to achieve more efficient algorithms \cite{varjas2020Local_marker_KPM}, though this approach's efficiency gains decrease as the bulk spectral gap decreases. Finally, none come equipped with an independent measure of the system's topological protection, and instead default to defining this protection in terms of the width of the relevant spectral gap.

\section{Introduction to the Spectral Localizer Framework \label{sec:SLintro}}

\begin{figure}[t]
    \centering
    \includegraphics[width=\columnwidth]{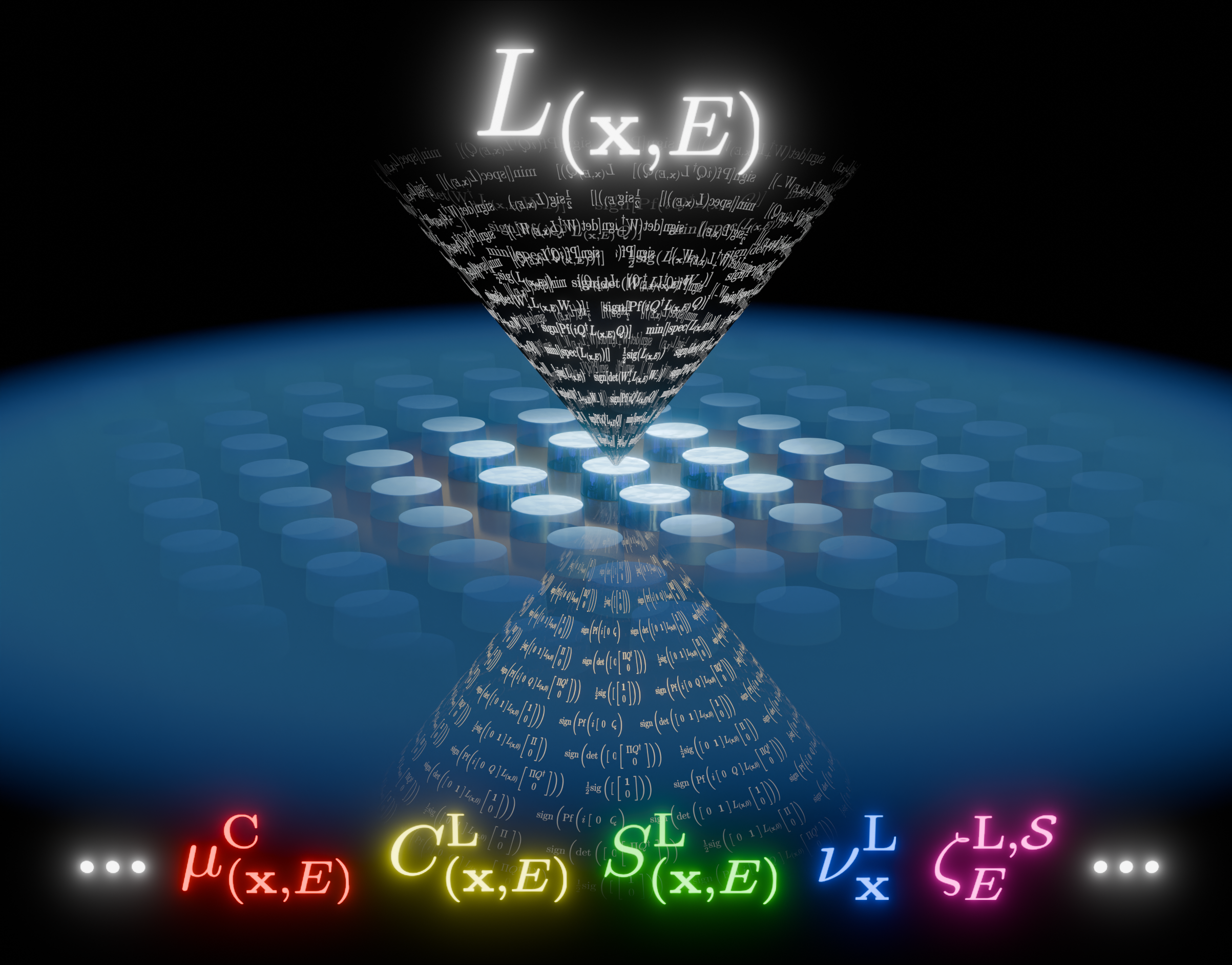}
    \caption{Schematic depiction of the spectral localizer probing a photonic metasurface at some location and outputting information about the material's properties at that location.}
    \label{fig:scheme}
\end{figure}

The spectral localizer framework can be thought of as a mathematical probe for understanding a material's properties near the probe's location, see Fig.~\ref{fig:scheme}. To classify material topology, the framework yields a constellation of local markers that can identify a broad range of material topology, including all ten Altland-Zirnbauer classes in every physical dimension \cite{loringPseuspectra,LoringSchuBa_odd,LoringSchuBa_even}, crystalline topology \cite{cerjan_local_2024}, Weyl semimetals \cite{schulz-baldes_spectral_2023}, and some forms of non-Hermitian topology \cite{liu_mixed_2023,ochkan_non-hermitian_2024,chadha_real-space_2024}. In addition, the framework provides a quantitative measure of topological protection. 
In contrast to other theories of local markers, the spectral localizer framework's markers are guaranteed to be integer valued, though the theory still requires a choice of hyper-parameter that plays a similar role to the choice of integration area required for the Kitaev and Bianco-Resta markers (see Sec.~\ref{sec:kappa}). Computationally, a key property of the spectral localizer framework is that it does not require finding a system's eigenstates, nor make use of a system's projector onto an occupied subspace, and thus can leverage significant numerical speedups using matrix factorization techniques (see Sec.~\ref{sec:numK}).
The theory was originally discovered by Loring in 2015 \cite{loringPseuspectra}, with substantial subsequent mathematical developments by Loring and Schulz-Baldes \cite{LoringSchuBa_odd,LoringSchuBa_even}.

In the remainder of this section, we first provide an argument for how the spectral localizer framework classifies Chern topology for 2D materials based on an analysis of the position-space behavior of atomic limits, and discuss how a measure of topological protection naturally appears. In Sec.~\ref{sec:genSL}, we provide the general definition of the spectral localizer, and discuss how it can classify topology in other symmetry classes. Later sections give further discussion on how the spectral localizer framework is applied to odd-dimensional systems, as well as how the framework can be applied to classify crystalline topology and higher-order topology. Finally, Sec.~\ref{sec:thermo} examines how the spectral localizer framework is generalized to the thermodynamic limit.

\subsection{Intuitive picture of the Spectral Localizer framework for identifying 2D Chern materials \label{sec:2dSLintro}}

Before proceeding to a more formal introduction, here we begin by introducing the spectral localizer framework from a ground-up perspective for identifying Chern topology in 2D materials. The goal is to provide a complete intuitive picture of this framework, while pushing some details to later sections. Although there are different ways to understand why the spectral localizer framework can successfully classify material topology, see Sec.~\ref{sec:whyWorks}, this section introduces the spectral localizer framework as a method for diagnosing whether a given system can be connected to an atomic limit without closing a spectral gap or violating a relevant symmetry.

An important concept in the modern study of topological materials is the idea of an \textit{atomic limit} --- the limit in which the constituent elements of a material are decoupled into individual atoms, molecules, or meta-atoms so that the system is simply a collection of those isolated elements. A crystal in an atomic limit exhibits a band structure that is completely flat \cite{kitaev2009}. Atomic limits also possess a complete basis of localized Wannier functions \cite{marzari_maximally_2012} that exhibit all of the same symmetries as the underlying system. Atomic limits are intimately connected to material topology because topologically non-trivial systems either do not possess a localized Wannier basis, or their localized Wannier basis does not obey all of the symmetries of the original material. For example, 2D systems with non-zero Chern numbers do not possess a localized Wannier basis \cite{brouder_exponential_2007}, which, in crystalline insulators, is a direct consequence of the fact that such systems have an obstruction that prohibits the choice of a smooth gauge for the Bloch wavefunctions across the (first) Brillouin zone. Similarly, electronic systems that exhibit the quantum spin Hall effect, a form of topology protected by fermionic time-reversal symmetry, possess a localized Wannier basis, but this basis does not obey fermionic time-reversal symmetry \cite{soluyanov_wannier_2011}.

\begin{figure*}[t]
    \centering
    \includegraphics{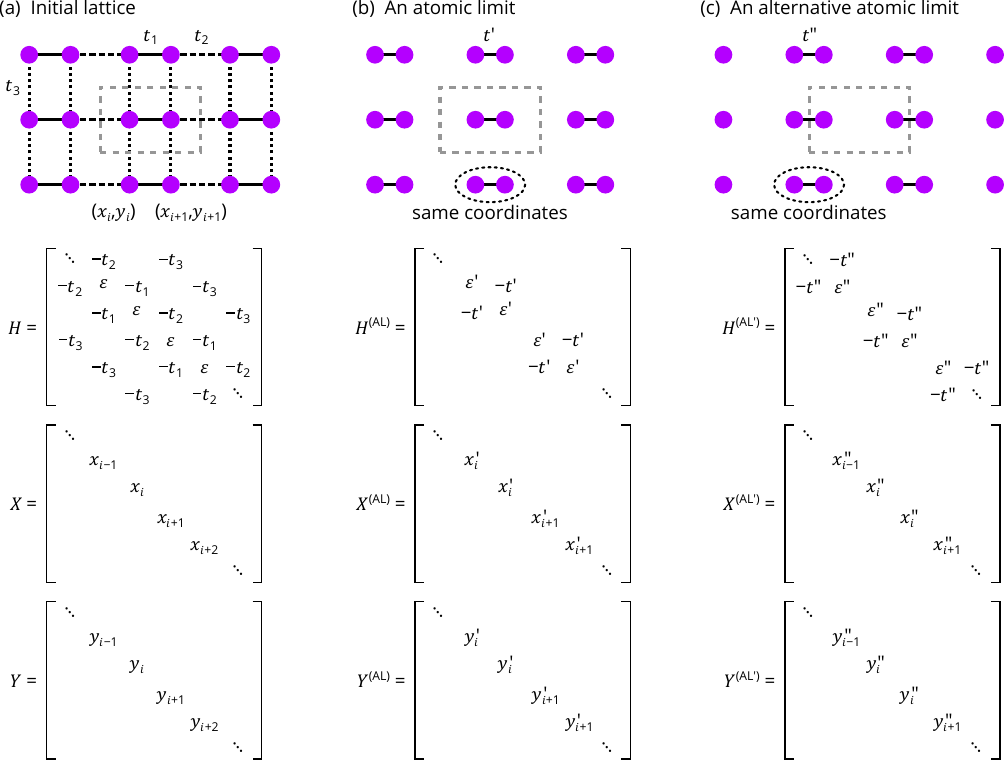}
    \caption{Schematic representation of a 2D tight-binding lattice with rectangular symmetry (a), as well as two of the possible atomic limits that this system might be connected to (b,c). For each lattice, we illustrate the form that their tight-binding Hamiltonian and associated position operators take.}
    \label{fig:AL}
\end{figure*}

Crucially, these statements are bijections, e.g., if a 2D material is found to not possess a localized Wannier basis, it necessarily has a non-zero Chern number. If a material does not possess a localized Wannier basis that exhibits all of the same symmetries as the original system, the system is generally non-trivial with respect to a class of topology protected by those symmetries missing from the Wannier basis. In either case, the material cannot be continued to an atomic limit without breaking a symmetry or closing a spectral gap (which would allow the system to change its topology). Here, by ``continued to an atomic limit,'' one is asking whether a path of matrices $H_\tau$ with $\tau \in [0,1]$ can be found with $H_0 = H$ being the original system's Hamiltonian and $H_1 = H^{(\textrm{AL})}$ being an atomic limit, such that every $H_\tau$ possesses a bulk spectral gap at the chosen energy $E$, remains somewhat local \cite{kitaev2009} (i.e., sites separated by a sufficiently large distance are not coupled), and obeys the same symmetries as $H$ (e.g., if a system is chiral symmetric with $H\Pi = -\Pi H$, then for every $\tau$, $H_\tau \Pi = -\Pi H_\tau$).
As such, it is possible to flip the paradigm for identifying topological systems: rather than finding materials, calculating their topological invariants, and inferring the existence and properties of the system's localized Wannier basis; one can instead choose a material, ask if it exhibits a localized Wannier basis that obeys all of the same symmetries as the original material, and infer its topology. Indeed, the recently developed framework for classifying crystalline materials of topological quantum chemistry \cite{kruthoff_topological_2017,bradlyn_topological_2017,po_symmetry-based_2017,cano_building_2018,watanabe_space_2018,de_paz_engineering_2019,elcoro_magnetic_2021,christensen_location_2022} takes precisely this approach, using a material's band structure to understand whether it can be continued to an atomic limit and then inferring its topological properties.

However, it is also possible to ascertain whether a system can be continued to an atomic limit directly from position-space description, rather than through its band structure or Wannier basis. The key mathematical observation for this shift in perspective follows from the definition of an atomic limit: since all of an atomic limit's constituent atoms or molecules are decoupled, the system does not possess any kinetic energy associated with this decoration-to-decoration coupling; moreover, the spacing between adjacent atoms or molecules is assumed to be large compared with the spacing between the elements within a single molecule. Thus, the Hamiltonian of an atomic limit commutes with its associated position operators $X_j^{(\textrm{AL})}$, i.e., $[H^{(\textrm{AL})},X_j^{(\textrm{AL})}]=0$ $\forall j$, see Fig.~\ref{fig:AL}. As such, from a position-space perspective, the question of whether a given system can be continued to an atomic limit is equivalent to asking whether the non-commuting Hamiltonian $H$ and position operators $X_j$ of the original system, $[H,X_j] \ne 0$, can nevertheless be path continued via some set of $H_\tau$ and $X_{j,\tau}$ to be commuting while preserving all of the relevant symmetries and the relevant bulk spectral gap. %In other words, is there a path of $H_\tau$ and $X_{j,\tau}$ with $H_0 = H$, $X_{j,0} = X_j$, $H_1 = H^{(\textrm{AL})}$, and $X_{j,1} = X_j^{(\textrm{AL})}$
Note that in this position-space picture, the effect of symmetries must be considered on both $H_\tau$ and $X_{j,\tau}$; while local symmetries like time-reversal trivially commute with $X_{j,\tau}$, crystalline symmetries might not, and the path of $X_{j,\tau}$ must preserve whatever the original relationship is. Additionally, at this juncture we are being purposefully vague about how to guarantee the position-space path preserves the relevant bulk spectral gap; for a finite system with OBC, edge effects, topological or otherwise, as well as internal defects may result in $H$ not exhibiting any global spectral gap even if its ordered crystalline counterpart exhibits bulk band gaps. Exactly what spectral gap must be preserved by $H_\tau$ and $X_{j,\tau}$ will be made rigorous in Sec.~\ref{sec:mu}.

\begin{figure}[t]
    \centering
    \includegraphics{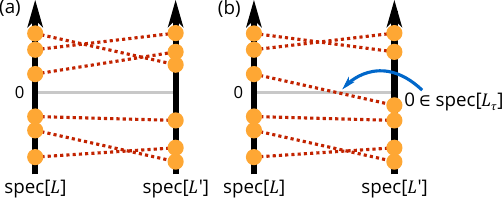}
    \caption{(a) Schematic showing the how $L$ and $L'$ belong to the same homotopy class because they have the same signature and can thus be connected by a path of invertible matrices $\{L_\tau\}$. An example homotopy-preserving path is shown by the red dotted lines. (b) Similar, but now $\textrm{sig}[L] \ne \textrm{sig}[L']$, so any attempt to construct a path of Hermitian matrices $\{L_\tau\}$ to connect them must become non-invertible somewhere, i.e., there will be some $\tau$ for which $0 \in \spec[L_\tau]$. Here, $\spec[M]$ is the spectrum of $M$.}
    \label{fig:sigThm}
\end{figure}

Having reduced the question of classifying material topology to determining whether a set of non-commuting matrices can be appropriately path-continued to be commuting, a method must be argued for actually performing this determination. For a 2D Chern insulator, we can understand whether a given material can be path continued to the atomic limit using two theorems from the mathematics literature. The first theorem provides a way to identify the homotopy class of an invertible Hermitian matrix, i.e., it provides a method for identifying whether two invertible Hermitian matrices can be path connected.
\vspace{5px}
\newline \noindent \textbf{Theorem A:} Two $n$-by-$n$ invertible Hermitian matrices $L$ and $L'$ can be connected by a path of invertible Hermitian matrices if and only if $\textrm{sig}[L] = \textrm{sig}[L']$, where $\textrm{sig}[L]$ is the \textit{signature} of $L$, its number of positive eigenvalues minus its number of negative eigenvalues. (See App.~\ref{app:homotopy} for a proof.)
\vspace{5px}
\newline \noindent While this theorem uses language that is not standard for photonics, the result is quite intuitive when depicted graphically: Any attempt to connect the spectra of two invertible Hermitian matrices with different signatures will necessarily force at least one eigenvalue in the connecting path of Hermitian matrices $L_\tau$ to be $0$, at which point that $L_\tau$ is not invertible, see Fig.~\ref{fig:sigThm}. With knowledge of what matrices can be connected, the second theorem then defines which of these homotopy classes contain atomic limits.
\vspace{5px}
\newline \noindent \textbf{Theorem:} (Choi, 1988 \cite{choi_almost_1988}, Lemma 4) Given two $N$-by-$N$ Hermitian matrices $R$ and $S$, if $[R,S] = 0$, then
\begin{equation}
    \textrm{sig}\left[\begin{array}{cc}
    R & S \\
    S^\dagger & -R
    \end{array} \right] = 0.
\end{equation}
%\vspace{5px}
%\newline
\noindent While it is not immediately obvious in Choi's notation how this theorem is relevant, a 2D system's physical behavior can be recovered by substituting $R \rightarrow H$ and $S \rightarrow \kappa (X - iY)$, where $\kappa > 0$ is a scaling coefficient that ensures consistent units. (A complete discussion of $\kappa$ is provided in Sec.~\ref{sec:kappa}.) Then, the theorem's requirement that $R$ and $S$ commute becomes a requirement on the Hamiltonian commuting with the position matrices. In other words, Choi's theorem states that for an atomic limit
\begin{equation}
    \textrm{sig}\left[\begin{array}{cc}
        H^{(\textrm{AL})} & \kappa(X^{(\textrm{AL})} - iY^{(\textrm{AL})}) \\
        \kappa(X^{(\textrm{AL})} + iY^{(\textrm{AL})}) & -H^{(\textrm{AL})}
    \end{array} \right] = 0. \label{eq:choiAL}
\end{equation}

As Choi's theorem guarantees that a specific invertible Hermitian matrix comprised of an atomic limit's Hamiltonian and position operators will have zero signature, and the first theorem tells us that invertible Hermitian matrices with the same signature have the potential to be path-connected, together, they suggest an approach to classifying topology using a system's position-space description. Through analogy with \eqref{eq:choiAL}, the 2D spectral localizer is defined as
\begin{multline}
    L_{(\vec{x},E)}^{(\textrm{2D})}(X,Y,H) \\
    = 
    \left[ \begin{array}{cc}
    H-E\mathbf{1} & \kappa(X-x\mathbf{1}) - i\kappa (Y-y\mathbf{1}) \\
    \kappa(X-x\mathbf{1}) + i \kappa(Y-y\mathbf{1}) & -(H-E\mathbf{1}) 
    \end{array} \right]. \label{eq:2dSL}
\end{multline}
Relative to \eqref{eq:choiAL}, \eqref{eq:2dSL} shifts the system's energy and position spectra to be centered around a chosen $(\vec{x},E)$. Mathematically, these offsets do not change the behavior of atomic limits, if $[A,B] = 0$, then $[A-a\mathbf{1},B-b\mathbf{1}] = 0$. Physically, re-centering the system's operators to $(\vec{x},E)$ is necessary because different physical or spectral locations in a system can exhibit different topology. Indeed, Choi's requirement that the matrices be finite means that $H$, $X$, and $Y$ in \eqref{eq:2dSL} describe a finite system with open boundaries and bounded energy spectra, so the resulting framework must be sensitive to these physical boundaries and the choice of energy. (Note, at this point, we have not provided a sufficient argument to discount the possibility of Choi's theorem applying to finite systems with PBC, but this possibility is precluded by the need to establish a bulk-boundary correspondence for the spectral localizer framework, see Sec.~\ref{sec:bbc}.)

Finally, a system's local Chern marker can be defined using the 2D spectral localizer as
\begin{equation}
    C_{(\vec{x},E)}^{\textrm{L}}(X,Y,H) = \frac{1}{2}\textrm{sig} \left[ L_{(\vec{x},E)}^{(\textrm{2D})}(X,Y,H)\right] \in \mathbb{Z}. \label{eq:SLchern}
\end{equation}
In other words, if at some choice $(\vec{x},E)$ the 2D spectral localizer has a zero signature, the system's non-commuting $H-E\mathbf{1}$ and $X-x\mathbf{1}$, $Y-y\mathbf{1}$ can nevertheless be locally path continued to be commuting, i.e., to be in an atomic limit. As no symmetries have been specified, \eqref{eq:SLchern} is classifying material topology without respect to any symmetries, i.e., whether it is a Chern material (2D Class A).  Moreover, for semi-infinite crystalline systems with bulk band gaps, $C_{(\vec{x},E)}^{\textrm{L}}$ is provably equal to the global Chern number at the same energy, $C_{E}$ from \eqref{eq:bandC} \cite{LoringSchuBa_even} (up to a sign ambiguity, see Sec.~\ref{sec:genSL}). Finally, in \eqref{eq:SLchern}, the factor of $1/2$ corrects for the fact that when one eigenvalue of a matrix switches sign, the matrix's signature changes by $2$.

Overall, the argument presented in this section shows how the 2D spectral localizer can be used to classify whether a material at a given choice of $(\vec{x},E)$ exhibits non-trivial local Chern topology by understanding if a that system can locally be path continued to an atomic limit. In addition, the calculation of a system's local Chern marker can be computationally quite efficient, and need not involve finding any eigenvalues of $L_{(\vec{x},E)}^{(\textrm{2D})}$ by taking advantage of sparse matrix factorization techniques, see Sec.~\ref{sec:numK}.

\subsection{Defining the local gap \label{sec:mu}}

In Sec.~\ref{sec:2dSLintro}, we argued that for a system to be topologically trivial, the path of matrices $H_\tau$ and $X_{j,\tau}$ connecting a given system to an atomic limit need to both respect all of the original system's (relevant) symmetries and maintain a spectral gap. Then, we stated that the 2D spectral localizer must be applied to finite systems with OBC. However, this seems to present a problem: the Hamiltonian of a topological system with OBC will not generally exhibit a spectral gap, such a gap will instead by populated by boundary-localized states. Moreover, even topologically trivial systems may exhibit edge effects or possess internal defects that preclude its Hamiltonian from exhibiting a spectral gap. 

Instead, to see what gap must be preserved by the path of matrices $H_\tau$ and $X_{j,\tau}$, note that, by definition, $C_{(\vec{x},E)}^{\textrm{L}}$ cannot change its value unless at least one of the eigenvalues of the spectral localizer crosses $0$. Thus, consider the absolute value of the eigenvalue of $L_{(\vec{x},E)}$ that is closest to zero,
\begin{equation}
    \mu_{(\vec{x},E)}^{\textrm{C}}(\vec{X},H) = \min \left[ \left| \spec \left(L_{(\vec{x},E)}(\vec{X},H) \right) \right| \right], \label{eq:muC}
\end{equation}
where $\vec{X} = (X_1,...,X_d)$ for a $d$ dimensional system and $\spec[M]$ is the spectrum of $M$. Here, the superscript $\textrm{C}$ denotes Clifford, as $\{ (\vec{x},E) | \mu_{(\vec{x},E)}^{\textrm{C}} \le \epsilon \}$ defines the Clifford $\epsilon$-pseudospectrum of $(\vec{X},H)$, see Sec.~\ref{sec:moPseudo}. Note that, as $L_{(\vec{x},E)}$ has units of energy, so does $\mu_{(\vec{x},E)}^{\textrm{C}}$. So long as $\mu_{(\vec{x},E)}^{\textrm{C}}(\vec{X}_\tau,H_\tau) > 0$ along the entire path $\tau$, the local Chern marker at $(\vec{x},E)$ cannot change. As such, the spectral gap that must be preserved to understand whether a system can be path connected to an atomic limit is that of $L_{(\vec{x},E)}(\vec{X}_\tau,H_\tau)$, not $H_\tau$ or $X_{j,\tau}$ individually.

However, the argument that a system's local topology is preserved so long as $\mu_{(\vec{x},E)}^{\textrm{C}} > 0$ has much broader implications for the spectral localizer framework: it defines a local measure of topological protection. As $L_{(\vec{x},E)}(\vec{X},H)$ is Hermitian, its eigenvalues $\lambda_j^{(H)}$ are guaranteed to be real, which means that they can be ordered, such that $\lambda_{j-1}^{(H)} \le \lambda_j^{(H)}$ $\forall j$. The same is true of the spectral localizer $L_{(\vec{x},E)}(\vec{X},H + \delta H)$ for the system with some perturbation $\delta H$, i.e., $\lambda_{j-1}^{(H+\delta H)} \le \lambda_j^{(H + \delta H)}$ $\forall j$. As the eigenvalues of each Hermitian matrix are ordered and can be paired up, we can find the distance each has moved for any perturbation, and Weyl's inequality guarantees that this distance is bounded by the norm of the difference in the operators \cite{weyl_asymptotische_1912,Bhatia1997},
\begin{equation}
    |\lambda_j^{(H + \delta H)} - \lambda_j^{(H)}| \le \Vert L_{(\vec{x},E)}(\vec{X},H + \delta H) - L_{(\vec{x},E)}(\vec{X},H) \Vert. \label{eq:weyl}
\end{equation}
Here, $\Vert M \Vert$ denotes the largest singular value of $M$, i.e., the $L^2$ matrix norm. In other words, the spectrum of the spectral localizer is Lipschitz continuous with coefficient $1$, see Ref.~\citen{loringPseuspectra}, Lemma 7.1. Moreover, due to the structure of the spectral localizer and the fact that the perturbation is only in the system's Hamiltonian,
\begin{equation}
    \Vert L_{(\vec{x},E)}(\vec{X},H + \delta H) - L_{(\vec{x},E)}(\vec{X},H) \Vert = \Vert \delta H \Vert.
\end{equation}

Now, let the perturbation be strong enough to reach a local topological phase transition, i.e., $\mu_{(\vec{x},E)}^{\textrm{C}}(\vec{X},H + \delta H) = 0$. As the eigenvalues of the spectral localizer are ordered, the shortest distance that an eigenvalue of the original system needs to move to reach such a phase transition point is $\mu_{(\vec{x},E)}^{\textrm{C}}(\vec{X},H)$. But, by \eqref{eq:weyl}, this means that the strength of the perturbation necessary to reach a phase transition point has a lower bound,
\begin{equation}
    \Vert \delta H \Vert \ge \mu_{(\vec{x},E)}^{\textrm{C}}(\vec{X},H).
\end{equation}
Thus, any perturbation with norm less than $\mu_{(\vec{x},E)}^{\textrm{C}}(\vec{X},H)$ cannot change the system's local topology. Altogether, this means that $\mu_{(\vec{x},E)}^{\textrm{C}}$ provides a rigorous definition of a \textit{local gap}; large values of $\mu_{(\vec{x},E)}^{\textrm{C}}$ relative to the energy scale of a perturbation
%quantities like $([\kappa X_j, H])^{(1/2)}$
guarantee that any local topological phase is robust, see Sec.~\ref{sec:moPseudo}. Finally, a similar argument shows that the size of $\mu_{(\vec{x},E)}^{\textrm{C}}$ also puts a bound on how far away one needs to look in $(\vec{x},E)$-space to have the possibility of a local change in topology; as changes in $(\vec{x},E)$ can only potentially close $\mu_{(\vec{x},E)}^{\textrm{C}}$ so quickly.

Generally, when using the spectral localizer to classify the topology of a periodic systems with a bulk band gap, the topological protection predicted by $\mu_{(\vec{x},E)}^{\textrm{C}}$ for $\vec{x}$ chosen in a system's bulk and $E$ in its spectral gap is quantitatively similar to the topological protection predicted by its band gap. \cite{cerjan_operator_Maxwell_2022}. In particular, if a system's bulk spectral gap is between a lower energy $E_{\textrm{l}}$ and an upper energy $E_{\textrm{u}}$, $\mu_{(\vec{x},E)}^{\textrm{C}}$ is typically within $\sim 10\%$ of $\min [E-E_{\textrm{l}}, E_{\textrm{u}}-E]$. However, for systems or heterostructures lacking bulk spectral gaps, the spectral localizer framework can still predict non-zero values for topological protection \cite{cerjan_local_2022,cheng_revealing_2023,dixon_classifying_2023}.

\begin{figure}[t]
    \centering
    \includegraphics{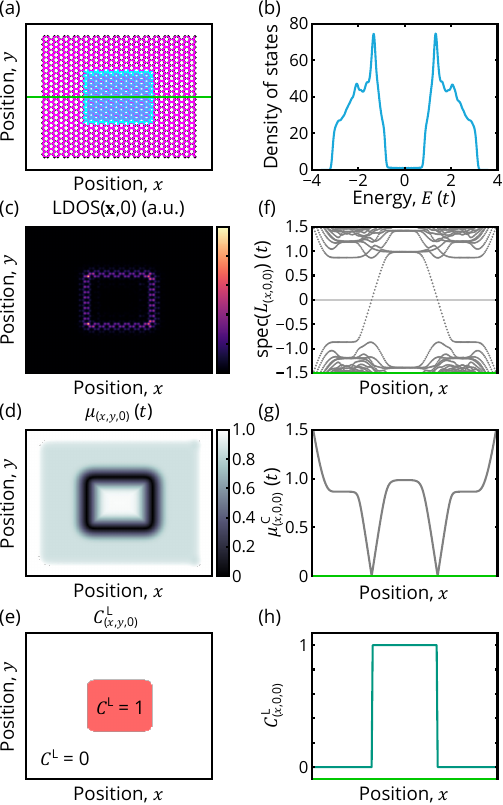}
    \caption{(a) Schematic of a heterostructure consisting of a Haldane lattice embedded within a massive graphene lattice. Both lattices have a nearest neighbor coupling strength of $t$. The Haldane lattice is massless and has directional next-nearest neighbor couplings $t_2 e^{\pm i \phi}$ with $t_2 = 0.5t$ and $\phi = \pi/2$. The massive graphene lattice has on-site energies $\pm M$ with $M = t$. (b) Density of states of this finite heterostructure. (c,d,e) LDOS (c), local gap (d), and local Chern marker (e) all shown at $E = 0$ and on the same spatial scale as (a). (f,g,j) Spectrum of the spectral localizer (f), local gap (g), and local Chern marker (h) at $E=0$ as $x$ is varied along the green line path shown in (a). Simulations use $\kappa = 0.25(t/a)$ where $a$ is the site-to-site spacing. For the LDOS, each lattice site is represented as a 2D Gaussian with width $r = 0.3a$.}
    \label{fig:haldane}
\end{figure}

To summarize the picture of the spectral localizer framework, for a given system defined by $H$ and $\vec{X}$, one can first construct the spectral localizer, and then iterate over different choices of position and energy to map out the system's local topology and corresponding topological protection. As $L_{(\vec{x},E)}$ is usually sparse, this iterative process can be made quite efficient, see Sec.~\ref{sec:numK} for more details. An example appliation of the spectral localizer framework to a tight-binding topological heterostructure is shown in Fig.~\ref{fig:haldane}. Here, a finite region of a Haldane lattice \cite{haldane_model_1988} in a topological phase is embedded in a trivial insulator, and ultimately surrounded by open boundaries. The density of states and local density of states (LDOS) confirm that the heterostructure exhibits bulk spectral gap that is populated by chiral edge states localized to the interface between the two constituent materials, see Fig.~\ref{fig:haldane}b,c. Furthermore, at the material interface, the spectral localizer framework shows that the local gap $\mu_{(\vec{x},E)}^{\textrm{C}} \rightarrow 0$ so that the local Chern marker can change its value and identify the inner material as topological, see Fig.~\ref{fig:haldane}d,e. The spectrum of the spectral localizer $\spec [L_{(\vec{x},E)}]$ shows a single eigenvalue crossing $0$ that captures the change in the system's local topology as $(\vec{x},E)$ is varied across the heterostructure with $E$ in the bulk spectral gap, see Fig.~\ref{fig:haldane}f. This flow of the eigenvalue is called the \textit{spectral flow} of $L_{(\vec{x},E)}$. Finally, this example also demonstrates that $(\vec{x},E)$ can be chosen to be any real numbers, with the edges of the plots showing values for $\vec{x}$ outside of the heterostructure.

\subsection{General definition of the spectral localizer in arbitrary dimensions \label{sec:genSL}}

Having demonstrated how the spectral localizer framework can be applied to classify 2D Chern materials, we now generalize the definition of the spectral localizer to an arbitrary number of physical dimensions, and briefly discuss the structure of the other local markers for classifying other forms of topology that can be defined using $L_{(\vec{x},E)}$. To do so, first note that for a 2D system, \eqref{eq:2dSL} can be rewritten using the Kronecker product and the Pauli matrices $\sigma_{x,y,z}$ as
\begin{multline}
    L_{(\vec{x},E)}^{(\textrm{2D})}(X,Y,H) \\
    = \kappa (X-x\mathbf{1}) \otimes \sigma_x + \kappa (Y-y\mathbf{1}) \otimes \sigma_y + (H-E\mathbf{1}) \otimes \sigma_z.
\end{multline}
This form hints at the correct structure of the spectral localizer for an arbitrary number of physical dimensions. Indeed, for a $d$-dimensional system, the spectral localizer is defined using a sufficiently large irreducible Clifford representation $\Gamma_j$ as
\begin{equation}
    L_{(\vec{x},E)}(\vec{X},H)
    = \sum_{j=1}^{d} \kappa (X_j - x_j \mathbf{1}) \otimes \Gamma_j + (H - E\mathbf{1}) \otimes \Gamma_{d+1}. \label{eq:genSL}
\end{equation}
To form a Clifford representation, $\Gamma_j$ must satisfy the following set of relations,
\begin{subequations} \label{eq:Clifford_Relations}
\begin{align}
\Gamma_j \Gamma_l & = - \Gamma_l \Gamma_j , \\ 
\Gamma_j^2 & = \mathbf{1}, \\  
\Gamma_j &= \Gamma_j^\dagger.
\end{align}    
\end{subequations}
where these must hold for all $1\leq j, l \leq d+1$ with $j\neq l$ in the first equation.
Irreduciblity is defined to mean that these $\Gamma_j$ are not built by stacking smaller Clifford representations. This precludes choosing a Clifford representation such as 
$\big[\begin{smallmatrix}\sigma_{x} & 0\\
0 & -\sigma_{x}
\end{smallmatrix}\big]$, $\big[\begin{smallmatrix}\sigma_{y} & 0\\
0 & -\sigma_{y}
\end{smallmatrix}\big]$, and $\big[\begin{smallmatrix}\sigma_{z} & 0\\
0 & -\sigma_{z}
\end{smallmatrix}\big]$ that would yield a value of zero in the local Chern marker formula
\eqref{eq:SLchern}.  Fortunately, we need only check that the $\Gamma_j$ have size $2^{\lceil d/2 \rceil}$ to guarantee irreducibility \cite{cerjan_even_2024} (where $\lceil d/2 \rceil$ denotes the ceiling of $d/2$).
As an example for a higher-dimensional system, the 4D spectral localizer can be constructed using $\Gamma_1 = \sigma_x \otimes \sigma_x$, $\Gamma_2 =  \sigma_y \otimes \sigma_x $, $\Gamma_3 = \sigma_z \otimes \sigma_x$, $\Gamma_4 =  \mathbf{1} \otimes\sigma_y $, and $\Gamma_5 = \mathbf{1} \otimes \sigma_z$, which are an irreducible Clifford representation for $d=4$. The first four of these are an irreducible Clifford representation for $d=3$.
(Note, the four Dirac matrices do not form a representation of Eqs.~\ref{eq:Clifford_Relations} since one of the Dirac matrices squares to $-\mathbf{1}$.)

Intuitively, the use of a Clifford representation in the spectral localizer framework can be understood from the need to preserve the ``orthogonality'' of the information present in the system's Hamiltonian and position operators while still combining these operators into a single $L_{(\vec{x},E)}$. This is akin to how the Pauli matrices form a complete basis for all $2$-by-$2$ Hermitian matrices with spectra centered at $0$.
Rigorously, underlying the need for a Clifford representation in the spectral localizer framework lurks Clifford algebras, see Sec.~\ref{sec:whyWorks}. 

The spectral localizer framework provides a complete set of local markers for classifying any form of topology from the Altland-Zirnbauer classes in any physical dimension. In Sec.~\ref{sec:2dSLintro} we discussed how $\textrm{sig}[L_{(\vec{x},E)}^{(\textrm{2D})}]$ identifies the underlying system's Chern topology as this local marker does not require specifying any system symmetry. Similarly, the 2nd Chern number of a 4D system in Class A or AI can be found using a related invariant \cite{cerjan_even_2024},
\begin{equation}
    \frac{1}{2}\textrm{sig} \left[ L_{(\vec{x},E)}^{(\textrm{4D})}(X_1,X_2,X_3,X_4,H)\right] \in \mathbb{Z}.
\end{equation}
The spectral localizer can also identify topology that falls outside of the Altland-Zirnbauer classification, such as material topology associated with crystalline symmetries \cite{cerjan_local_2024}, see Sec.~\ref{sec:crystal}, as well as Weyl materials \cite{schulz-baldes_spectral_2023} and some forms of non-Hermitian topology \cite{liu_mixed_2023,ochkan_non-hermitian_2024,chadha_real-space_2024}.

Many of the real Altland-Zirnbauer classes are inaccessible in photonics, as they require local symmetries that do not appear in photonic systems such as fermionic time-reversal symmetry or particle-hole symmetry. Nevertheless, to illustrate the structure of these other local markers and demonstrate how they explicitly rely upon the symmetry protecting the topology, we will briefly review how the spectral localizer framework identifies the topology of two real Altland-Zirnbauer classes that exhibit $\mathbb{Z}_2$ invariants.
For example, the spectral localizer framework can identify quantum spin Hall materials (2D class AII), whose topology is protected by fermionic time-reversal symmetry $\mathcal{T}$, using the local marker \cite{loringPseuspectra}
\begin{equation}
    S_{(\vec{x},E)}^{\textrm{L}}(X,Y,H) 
    = \textrm{sign}\left( \textrm{Pf}\left[i U^\dagger L_{(\vec{x},E)}^{(\textrm{2D})'}(X,Y,H) U \right] \right) \in \mathbb{Z}_2,
\end{equation}
where
\begin{multline}
    L_{(\vec{x},E)}^{(\textrm{2D})'}(X,Y,H) \\
    = \kappa (X-x\mathbf{1}) \otimes \sigma_z + \kappa (Y-y\mathbf{1}) \otimes \sigma_x + (H-E\mathbf{1}) \otimes \sigma_y. \label{eq:2dSLp}
\end{multline}
Here, the local marker relies on the fact that $\mathcal{T}H = H \mathcal{T}$, which both guarantees that there is a basis in which $H$ is purely imaginary so that $L_{(\vec{x},E)}^{(\textrm{2D})'}$ is real, and that a unitary $U$ exists such that $i U^\dagger L_{(\vec{x},E)}^{(\textrm{2D})'} U$ is skew-symmetric with a well-defined Pfaffian. If the sign of this Pfaffian is positive, the material is locally trivial, if negative, the material is locally topological. (Rigorously, $S_{(\vec{x},E)}^{\textrm{L}} \in \{1,-1\} \cong \mathbb{Z}_2$, but we are abusing notation and will simply state $S_{(\vec{x},E)}^{\textrm{L}} \in \mathbb{Z}_2$ as the isomorphism is trivial.) From the perspective of the path continuation arguments discussed surrounding Thm.~A and Fig.~\ref{fig:sigThm} in Sec.~\ref{sec:2dSLintro}, the path of matrices $L_\tau$ connecting the unitarily transformed spectral localizers of two $\mathcal{T}$-symmetric systems must now remain skew-symmetric for all $\tau$. As such, the eigenvalues of $L_\tau$ can only reach or cross $0$ in pairs, so that $\textrm{sig}[L_\tau] = 0$ $\forall \tau$, but at these closings the Pfaffian of $L_\tau$ might change sign, in which case the path is not homotopy-preserving. Likewise, 2D class DIII systems also exhibit a $\mathbb{Z}_2$ classification, and such a system's local topology can be found via \cite{loringPseuspectra}
\begin{equation}
    \textrm{sign}\left( \textrm{det}\left[i W_+^\dagger L_{(\vec{x},E)}^{(\textrm{2D})'}(X,Y,H) W_- \right] \right) \in \mathbb{Z}_2,
\end{equation}
where $W_+$ and $W_-$ are again basis dependent and constructed through considering the system's local symmetries.

Overall, there are only three different forms of the local markers that are defined using the spectral localizer for the ten Altland-Zirnbauer classes,
\begin{subequations} \label{eq:markerForms}
\begin{align}
    \textrm{sig}[\ast] & \in \mathbb{Z} \\
    \textrm{sign}\left[ \textrm{Pf} (\ast) \right] & \in \mathbb{Z}_2 \\
    \textrm{sign}\left[ \textrm{det} (\ast) \right] & \in \mathbb{Z}_2 
\end{align}
\end{subequations}
where $\ast$ denotes some matrix based on the spectral localizer of appropriate dimension. Heuristically, the forms of these local markers can be understood through simple consideration of the possible structures of Hermitian matrices. As discussed in Sec.~\ref{sec:2dSLintro} and Fig.~\ref{fig:sigThm}, an invertible Hermitian matrix's signature is an invariant of homotopy. However, a Hermitian matrix may possess additional structure that both forces its signature to be zero, but also enables classification with respect to so-called \textit{secondary invariants} \cite{MR285033,freed2021atiyah,doll2021skew} that are only meaningful when the primary invariant, i.e., the matrix's signature, is trivial. For example, a Hermitian matrix might be purely imaginary, guaranteeing it to be skew-symmetric (i.e., $A^\trans = -A$), for which $\textrm{sign}\left[ \textrm{Pf} (\ast) \right]$ distinguishes different homotopy classes. Similarly, the Hermitian matrix might be real-symmetric (i.e., $A^\trans = A$) and anti-commute with a Hermitian unitary matrix (sometimes referred to as being odd with respect to a \textit{grading} operator), in which case $\textrm{sign}\left[ \textrm{det} (\ast) \right]$ applied to an off-diagonal corner of the Hermitian matrix distinguishes different homotopy classes. See App.~\ref{app:homotopy} for homotopy arguments for why these can be non-trivial topological invariants.

Moreover, regardless of the specific form of any of the local markers defined using the spectral localizer, all of these markers are fundamentally connected to the spectrum of the spectral localizer and cannot change their value without first closing the local gap, $\mu_{(\vec{x},E)}^{\textrm{C}} = 0$. Thus, \eqref{eq:muC} always provides the measure of topological protection for any class of topology in the spectral localizer framework. Additionally, it has been proven that changing the irreducible Clifford representation used in \eqref{eq:genSL} does not alter the local gap, see Ref.~\citen{cerjan_even_2024}, Lemma 1.2, e.g., $L_{(\vec{x},E)}^{(\textrm{2D})}$ and $L_{(\vec{x},E)}^{(\textrm{2D})'}$ both yield the same $\mu_{(\vec{x},E)}^{\textrm{C}}$.

Finally, we would like to be able to provide more guidance on how to pick the irreducible Clifford representation for \eqref{eq:genSL}, but this is currently an open topic. To illustrate the present difficulty, consider the 2D spectral localizer defined using $-\sigma_x$, $-\sigma_y$, and $-\sigma_z$ instead of their positive counterparts. The negative Pauli matrices still form an irreducible Clifford representation, but the spectrum of the resulting spectral localizer will be multiplied by $-1$ relative to the standard choice of 2D spectral localizer given by \eqref{eq:2dSL}, yielding a sign flip of the local Chern marker \eqref{eq:SLchern}. Thus, the spectral localizer framework always predicts the correct number of chiral edge modes and is internally consistent for a given choice of Clifford representation. However, the spectral localizer framework can have a sign ambiguity when compared against other topological frameworks. \underline{Open question:} Further study is necessary to understand if there is an argument for constructively fixing the sign ambiguity, i.e., without an appeal to other topological invariants.

\subsection{Identifying topology in 1D chiral symmetric systems using the spectral localizer framework \label{sec:SL1d}}

So far, this tutorial has only discussed the spectral localizer's application to even-dimensional systems. However, applying the spectral localizer framework to odd-dimensional systems requires a slight modification. The initial challenge is that an irreducible Clifford representation for odd $d$ is always a truncation of an irreducible Clfford represention for $d+1$. That extra matrix, $\Gamma_{d+2}$, anticommutes with the prior $\{\Gamma_{1},...,\Gamma_{d+1}\}$ and so the spectral localizer will be off-block diagonal in an appropriate basis. For example, for a 1D system we can generally use $\Gamma_{1}=\sigma_{x}$ and $\Gamma_{2}=\sigma_{y}$ and so the spectral localizer is off-block diagonal, 
\begin{multline}
    L_{(x,E)}^{(\textrm{1D})}(X,H) \\
    = 
    \left[ \begin{array}{cc}
    0 & \kappa(X-x\mathbf{1}) - i(H-E\mathbf{1}) \\
    \kappa(X-x\mathbf{1}) + i (H-E\mathbf{1}) & 0 
    \end{array} \right]. \label{eq:1dSL}
\end{multline}
The off-block diagonal form of spectral localizer reveals the fundamental challenge in odd dimensions: the two blocks contain essentially opposite spectral information about the system, e.g., the signature of \eqref{eq:1dSL} is always zero. To remove this duplicate-but-opposite information, the odd-dimensional local markers are defined in terms of only one of these blocks. However, each block has no particular structure, being neither Hermitian nor real. Instead, to recover a Hermitian or real matrix whose homotopy can be classified using constructions like Eqs.~(\ref{eq:markerForms}), the system must exhibit a unitary or anti-unitary symmetry that anti-commutes with at least one of $X_j$ or $H$. Indeed, this mathematical symmetry requirement is in agreement with the Altland-Zirnbauer classification, which identifies odd-dimensional Class A systems as always being trivial.

For example, consider a 1D chiral symmetric system such as the Su-Schrieffer-Heeger (SSH) model \cite{su_soliton_1980} (1D class AIII) that is described by a Hamiltonian and a single position operator. The local winding number for such a system is
\begin{equation}
    \nu_x^{\textrm{L}}(X,H) = \frac{1}{2}\textrm{sig} \left[ \tilde{L}_{(x,0)}^{(\textrm{1D})}(X,H) \right] \in \mathbb{Z}. \label{eq:SLwind}
\end{equation}
in which the associated \textit{symmetry reduced spectral localizer} is 
\begin{equation}
    \tilde{L}_{(x,E)}^{(\textrm{1D})}(X,H) = \left[\kappa(X-x\mathbf{1}) - i(H-E\mathbf{1})\right] \Pi. \label{eq:redSL}
\end{equation}
Here, $\Pi$ is the chiral symmetry matrix, $H\Pi = -\Pi H$ and $X \Pi = \Pi X$, and the local marker is only defined at $E = 0$ as is expected for this Altland-Zirnbauer class. Note that although $\tilde{L}_{(x,E)}^{(\textrm{1D})}$ is generally non-Hermitian, at $E=0$, $\tilde{L}_{(x,0)}^{(\textrm{1D})}$ is Hermitian, and thus has a well-defined signature. A proof showing how systems with non-trivial local winding numbers cannot be continued to an atomic limit while preserving chiral symmetry can be found in Ref.~\citen{cerjan_local_2024}, Section SII.

\begin{figure}[t]
    \centering
    \includegraphics{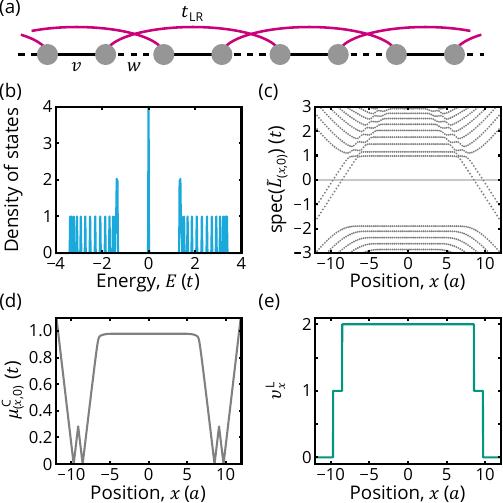}
    \caption{(a) Schematic of a SSH lattice with chiral-preserving long-range couplings, with $v = t$, $w = 0.4t$, and $t_{\textrm{LR}} = 2t$. Simulated system contains $20$ unit cells, with a lattice constant of $a$ and a site-to-site spacing of $0.5a$. The system terminates after a weak $w$ coupling on both ends. (b) Density of states for this system. (c,d,e) Spectrum of $\tilde{L}_{(x,0)}^{(\textrm{1D})} \Pi$ (c), local gap (d), and local winding number (e) as $x$ is varied across the lattice. Simulations use $\kappa = 0.5(t/a)$.}
    \label{fig:ssh}
\end{figure}

Altogether, through a judicious choice of elements of the Clifford representation used for an odd-dimensional system, all of the necessary spectral information for defining local topological markers is contained in a single off-diagonal block of the spectral localizer. As always, the topological protection associated with these local markers is still given by \eqref{eq:muC}, which is also equivalent to finding the smallest singular value of the reduced spectral localizer. An example of the spectral localizer framework applied to a SSH model with chiral-preserving long-range couplings is shown in Fig.~\ref{fig:ssh}. Here, the long-range couplings yield a local winding number of $2$ in the lattice's interior bulk, and the spectral flow of $\tilde{L}_{(x,0)}^{(\textrm{1D})}$ responsible for this change in the local index can be seen in Fig.~\ref{fig:ssh}c.

\subsection{Classifying crystalline topology \label{sec:crystal}}

A key benefit of the spectral localizer framework is that it is agnostic to the physical meaning of the matrices being used to construct it and its associated local markers. In particular, this means that if another system symmetry can be found that is outside of the standard Altland-Zirnbauer classes, and yet exhibits equivalent relations on the system's Hamiltonian and position operators, then the local marker for this symmetry-protected topology can be immediately constructed \cite{cerjan_local_2024}. 

For example, for 1D systems with chiral symmetry (1D class AIII), $H\Pi = -\Pi H$ and $X \Pi = \Pi X$, where $\Pi$ is the chiral operator; the associated local marker is given by \eqref{eq:SLwind}. Now, consider a crystalline symmetry $\mathcal{S}$ for which $H \mathcal{S} = \mathcal{S} H$ and $X \mathcal{S} = - \mathcal{S} X$, e.g., inversion symmetry. In such a case, the local topological marker can be easily constructed by swapping the Hamiltonian and position matrix $H \leftrightarrow X$ and the chiral operator with the crystalline symmetry $\Pi \leftrightarrow \mathcal{S}$ to define a local crystalline winding number
\begin{equation}
    \zeta_{E}^{\textrm{L},\mathcal{S}}(X,H) = \frac{1}{2}\textrm{sig} \left[ (H-E\mathbf{1} + i \kappa X) \mathcal{S} \right]. \label{eq:SLcrys}
\end{equation}
The associated measure of topological protection of is still defined by \eqref{eq:muC}.

There are two subtleties associated with the crystalline winding number. First, just as any topologically protected states in the SSH model must exist at $E=0$, the crystalline winding number identifies states at $x=0$, which is the center of the crystalline symmetry. Thus, this class of markers fixes $x$ and sweeps $E$ to identify material topology. Second, as the spectral localizer works with systems that possess OBC, the relevant crystalline symmetry $\mathcal{S}$ in \eqref{eq:SLcrys} is a global symmetry, not a single unit cell operator. \underline{Open question:} It remains an active area of research to explore whether the spectral localizer can be used to define topology with respect to single-unit-cell versions of a system's critical symmetries, rather than their global counterparts.

\begin{figure}[t]
    \centering
    \includegraphics{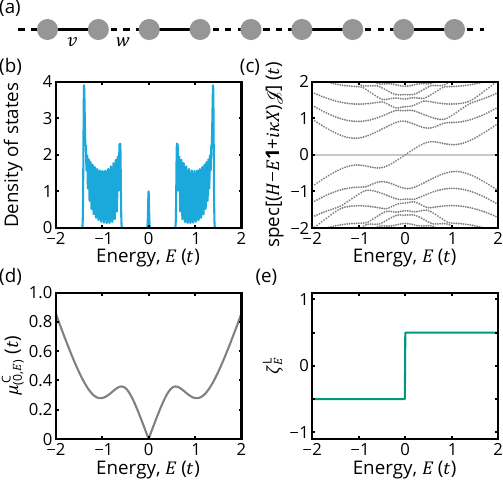}
    \caption{(a) Schematic of an inversion-symmetric SSH lattice with a defect site in the center, with $v = t$ and $w = 0.4t$. Simulated system contains $40$ unit cells and a single defect in the center (i.e., $81$ sites total) connected to the remainder of the lattice with weak $w$ couplings, with a lattice constant of $a$ and a site-to-site spacing of $0.5a$. The system terminates after a strong $v$ coupling on both ends. (b) Density of states for this system. (c,d,e) Spectrum of $(H-E\mathbf{1} + i \kappa X) \mathcal{S}$ (c), local gap (d), and local winding number (e) as $E$ is varied across the lattice's spectrum. Simulations use $\kappa = 0.5(t/a)$.}
    \label{fig:ssh_crys}
\end{figure}

An example of using the spectral localizer framework to classify crystalline topology is shown in Fig.~\ref{fig:ssh_crys}, where we consider an inversion-symmetric SSH model with a defect site in the center. Thus, for this system, $H \mathcal{I} = \mathcal{I} H$ and $X \mathcal{I} = -\mathcal{I} X$, where $\mathcal{I}$ is the inversion operator. The observation that $\zeta_{E}^{\textrm{L},\mathcal{I}}$ is taking half-integer values is not a mistake, but is instead a direct consequence of the fact that this system has an odd number of sites in conjunction with the $1/2$ in \eqref{eq:SLcrys}. These half-integers are not a problem, as differences in the local marker at different $E$ are still integer valued, the spectral flow of $(H-E\mathbf{1} + i \kappa X) \mathcal{I}$ across this range of $E$ is the same integer, and this integer corresponds to the number of inversion-center localized topological states exist. (See Sec.~\ref{sec:bbc} for a discussion of bulk-boundary correspondence.)

\subsection{Dimensional reduction and higher-order topology}

The spectral localizer framework can also dimensionally reduce a system and consider its topology in a lower dimension. As such, the framework can be used to identify higher-order topology \cite{benalcazar2017quad,benalcazar2017quadPRB,song_d-2-dimensional_2017,schindler_higher-order_2018,benalcazar_quantization_2019,benalcazar_chiral-symmetric_2022}, protected either by crystalline symmetry \cite{cerjan_local_2024} or by chiral symmetry \cite{cerjan_local_2022}. 
Mathematically, this dimensional reduction is achieved by simply omitting one or more position operators from the spectral localizer, and choosing a Clifford representation suitable for the effective lower-dimensional system. In doing so, the full system is still used, i.e., the Hamiltonian remains unchanged. Instead, this process is projecting the system into the lower-dimensional space and simply forgetting about their position in the omitted dimension(s). 
Physically, changes in a system's local topology can only occur at locations in position-energy space where $\mu_{(\vec{x},E)}^{\textrm{C}} = 0$, which also guarantees the system exhibits a nearby states, see Sec.~\ref{sec:bbc}. Thus, by projecting the system into a suitable lower-dimensional space, one can ensure that a path in $(\vec{x},E)$ taken by the spectral localizer always crosses a location where the local gap closes so the topology can change, precluding the possibility of choosing a path that goes around the local gap closure by omitting that dimension.

\begin{figure*}[t]
    \centering
    \includegraphics{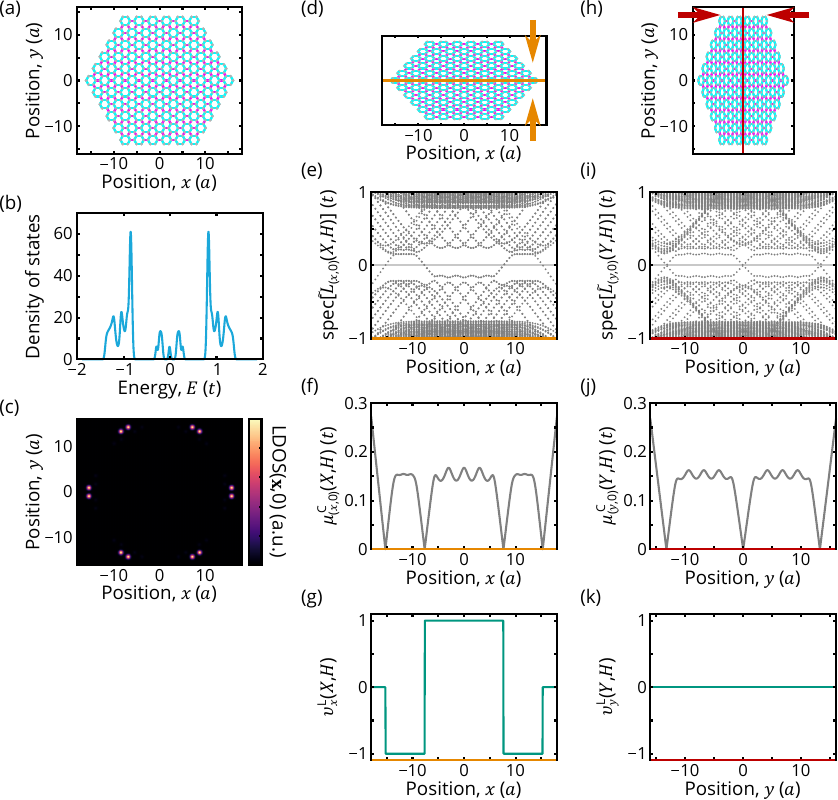}
    \caption{(a) Schematic of a 2D breathing honeycomb lattice with $C_{6v}$ and chiral symmetry. The intra-unit cell couplings have strength $t_{\textrm{in}} = 0.2t$, the inter-unit cell couplings have strength $t_{\textrm{out}} = t$, and the lattice constant is $a$. (b) Density of states for this system. (c) LDOS at $E=0$; each lattice site is represented as a 2D Gaussian with width $r = 0.3a$. (d,e,f,g) Application of the spectral localizer framework to the breathing honeycomb lattice dimensionally reduced to the $x$ axis, showing a schematic of the dimensional reduction (d), the spectral flow of $\tilde{L}_{(x,0)}^{(\textrm{1D})}(X,H)$ (e), the local gap $\mu_{(x,0)}^{\textrm{C}}(X,H)$ (f), and the local winding number $\nu_x^{\textrm{L}}(X,H)$ (g). (h,i,j,k) Are similar, but consider the same system dimensionally reduced to the $y$ axis, and report quantities based on $\tilde{L}_{(y,0)}^{(\textrm{1D})}(Y,H)$. Simulations use $\kappa = 0.1(t/a)$.}
    \label{fig:dimRed}
\end{figure*}

An example of using the spectral localizer to identify higher-order topology in a 2D breathing honeycomb lattice \cite{noh_topological_2018} is shown in Fig.~\ref{fig:dimRed}. The 2D breathing honeycomb lattice is chiral symmetric (class AIII), but in 2D this local symmetry class is always trivial. Instead, this lattice's higher-order topology and associated corner-localized states can be identified by projecting the lattice into a lower dimensional space and using the local winding number defined in \eqref{eq:SLwind}. In Fig.~\ref{fig:dimRed}d-g, the lattice is projected onto the $x$ axis. As $x$ is varied from negative to positive for $E=0$, the local winding number changes four times at positions that correspond to the projected locations of the system's six corners onto the $x$ axis. The first and last changes in $\nu_x^{\textrm{L}}(X,H)$ are by $\pm 1$, as these $x$ locations intersect a single lattice corner. The middle two changes in the local marker are by $\pm 2$, as the path is crossing two projected corners simultaneously. Crucially, because the corresponding topological corner-localized zero-energy modes have the same chiral charge, i.e., they predominantly have support on the same sublattice, their combined contribution to the local winding number is $\pm 2$. In contrast, projecting the lattice instead onto the $y$ axis yields a trivial local marker for every choice of $y$, see Fig.~\ref{fig:dimRed}h-k. Here, although the local gap closes three times where the six corners are projected onto the $y$ axis, each closing corresponds to the path crossing two corners with opposite chiral charge, yielding no change in $\nu_y^{\textrm{L}}(Y,H)$.

There is no requirement that the axis the system is projected onto for dimensional reduction is associated with one of the original position operators. For example, the 2D breathing honeycomb lattice in Fig.~\ref{fig:dimRed}a could be projected onto an arbitrary choice of axis $w = ax + by$ with associated position operator $W = aX + bY$. For most such choices, the associated local marker $\nu_w^{\textrm{L}}(W,H)$ would change six times, each by $\pm 1$.

Dimensional reduction plays an important role in the application of the spectral localizer framework to realistic photonic systems, as many technologically relevant photonic systems are 2.5D systems such as photonic crystal slabs and metasurfaces. Rigorously, such systems are 3D, but the desired topological boundary-localized states in these systems reside in the same planar slab as the structure. Thus, dimensional reduction enables the spectral localizer to rigorously define local markers associated with 2D material topology for such 3D planar systems. See Ref.\ \cite{wong_classifying_2024} for an example where the spectral localizer framework is used to identify a Chern photonic crystal slab.

\subsection{Behavior in the thermodynamic limit \label{sec:thermo}}

The structure of the spectral localizer's local markers presented in Sec.~\ref{sec:genSL} seems to suggest that these markers can only be applied to systems described by finite, but arbitrarily large, systems with bounded spectra where the signature, Pfaffian, and determinant operations are well-defined because, in principle, the entire spectrum of the spectral localizer can be found. (In practice, it is not recommended to ever compute the local markers this way, instead see Sec.~\ref{sec:numK}.)
Nevertheless, the definitions of the spectral localizer framework are still well-defined for some infinite-dimensional operators \cite{cerjan_multivariable_2024} such as those describing physical systems in the thermodynamic limit. The key is that formulas such as \eqref{eq:SLchern} can be generalized to infinite-dimensional $L_{(\vec{x},E)}$ by considering its spectral flow as $(\vec{x},E)$ moves along a path in position-energy space \cite{doll2023spectral}.

Consider a semi-infinite 2D system with a bounded spectrum. For $\vec{x}$ chosen far outside of the system, one can prove that the spectrum of its spectral localizer is balanced. Intuitively, this is straightforward to see, as if $\vec{x}$ moves away from the edge of the half lattice, $\kappa (X-x \mathbf{1})\otimes \sigma_x + \kappa (Y-y \mathbf{1})\otimes \sigma_y$ has a growing spectral gap that $(H-E\mathbf{1})\otimes \sigma_z$ is of too-small a norm to close. 
Heuristically, we want this to mean $\textrm{sig}[L_{(\vec{x},E)}^{(\textrm{2D})}] = 0$ as would be the case if the system were finite, but the signature is not defined for a semi-infinite system. To be rigorous, we instead need to use spectral flow and the so-called $\eta$-invariant \cite{doll2023spectral,LoringSchuBa_odd}. Then, as $\vec{x}$ is varied so that it crosses into the bulk of the system, the spectral localizer's spectrum near $0$ can be monitored for any crossings, i.e., one can track its spectral flow in an analogous manner to the examples shown in Figs.~\ref{fig:haldane}-\ref{fig:dimRed}. Crucially, despite the fact that the spectral localizer is infinite-dimensional (referring to the Hilbert space's dimension, not the number of physical dimensions), in many cases its spectral flow is well-defined and yields a local topological marker. In other words, in these cases spectral localizer has a ``signature'' that is twice the index of a related \textit{Fredholm operator} \cite{LoringSchuBa_odd,LoringSchuBa_even}. The $\mathbb{Z}_2$ indices relate to the secondary indices that can occur when the index of a Fredholm operator is zero \cite{doll2021skew}. Altogether, this means that the spectral localizer framework can be rigorously generalized to the thermodynamic limit without issue.

(At present, there is a technical restriction in defining the spectral flow of an infinite-dimensional operator, as we need to know that the spectral localizer has discrete spectrum, so we cannot now work directly with a differential operator such as those needed for photonic systems, see Sec.~\ref{sec:photon}. \underline{Open question:} We conjecture that the methods of spectral truncation \cite{connes2021spectral_truncation,vanSuijlekom2024spectral_truncation} in noncommutative geometry are likely to provide a way around this limitation.) 

\section{Underpinnings of the Spectral Localizer Framework \label{sec:4}}

Having introduced the spectral localizer framework and provided a number of examples of its application to identifying material topology across a variety of systems, this section discusses three topics that underpin how the spectral localizer framework functions. First, we provide a detailed discussion of the hyper-parameter $\kappa$ and provide some guidance for choosing $\kappa$. Then, we turn to introducing multi-operator pseudospectral methods, which form the foundation for proving bulk-boundary correspondence in the spectral localizer framework. Finally, we discuss how to efficiently implement the spectral localizer's local markers.

\subsection{The role of the hyper-parameter $\kappa$ \label{sec:kappa}}

\begin{figure*}[t]
    \centering
    \includegraphics{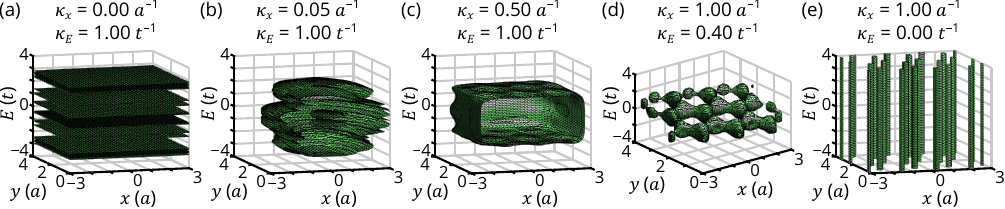}
    \caption{Evolution of the local gap closings as $\kappa$ is varied using a Haldane lattice. For this figure, the chosen lattice is very small, $6 \times 8$, to ease the computational requirement of finding these surfaces, so the corresponding range of $\kappa$ that yields the correct system topology is narrow. Calculations are performed using the dimensionless 2D spectral localizer \eqref{eq:dimlessSL}, and the surfaces correspond to $\mu_{(\vec{x},E)}^{\textrm{C}} = 0.05$ (a)-(d) and $\mu_{(\vec{x},E)}^{\textrm{C}} = 0.15$ (e).
    (a) $\kappa_x = 0$ and $\kappa_E = t^{-1}$.
    (b) $\kappa_x = 0.05 a^{-1}$ and $\kappa_E = t^{-1}$.
    (c) $\kappa_x = 0.5 a^{-1}$ and $\kappa_E = t^{-1}$.
    (d) $\kappa_x = a^{-1}$ and $\kappa_E = 0.4 t^{-1}$.
    (e) $\kappa_x = a^{-1}$ and $\kappa_E = 0$. Note, the choice of axis limits shown here bisects the Haldane lattice in $y$.}
    \label{fig:kappa}
\end{figure*}

The hyper-parameter $\kappa$ serves two critical roles in the spectral localizer framework. First, $\kappa$ adjusts the units of the position operators in $L_{(\vec{x},E)}$ to have dimensions of energy so that they can be combined with the Hamiltonian. Mathematically, this choice is arbitrary, the spectral localizer could be defined in units of length or made dimensionless through a second hyper-parameter and the structure of the local markers would not change. Physically, this choice is useful, as it enables direct comparison between the local gap $\mu_{(\vec{x},E)}^{\textrm{C}}$ and a system's bulk band gap or other spectral gap. Second, $\kappa$ balances the \textit{spectral weight} of the Hamiltonian relative to the position operators. In other words, $\kappa$ is chosen so that the eigenvalues of $L_{(\vec{x},E)}$ are similarly sensitive to changes in $H-E\mathbf{1}$ and changes in $\vec{X} - \vec{x}\mathbf{1}$, either because the choice of $(\vec{x},E)$ is shifted or because the system is perturbed $H \rightarrow H + \delta H$. Intuitively, this second function of $\kappa$ is playing a similar role as the choice of region areas in the Kitaev marker, \eqref{eq:kitaev}, or the choice of integration disk radius in the Bianco-Resta marker, \eqref{eq:bianco}. For example, larger values of $\kappa$ are comparable to smaller integration disk radii in the Bianco-Resta marker, and generally enable greater specificity in changes of $\vec{x}$ where the spectral localizer is evaluated, but potentially at the cost of being too insensitive to spectral information and mis-classifying the system's topology. Likewise, smaller values of $\kappa$ are comparable to larger integration disk radii, and generally yield correct material classification in the bulk until the corresponding length scale $\propto \kappa^{-1}$ is too large, similar to how the Bianco-Resta Chern marker is always trivial if the integration disk contains the entire lattice, see Fig.~\ref{fig:kappa}.

Fundamentally, it is somewhat remarkable that the spectral localizer can provide any useful information about a system at all. To illustrate this point, briefly consider the dimensionless 2D spectral localizer
\begin{multline}
    L_{(\vec{x},E)}^{(\textrm{2D})}(X,Y,H; \kappa_x, \kappa_E) \\
    = 
    \left[ \begin{array}{cc}
    \kappa_E(H-E\mathbf{1}) & \kappa_x(X-x\mathbf{1}) - i\kappa_x (Y-y\mathbf{1}) \\
    \kappa_x(X-x\mathbf{1}) + i \kappa_x(Y-y\mathbf{1}) & -\kappa_E (H-E\mathbf{1}) 
    \end{array} \right], \label{eq:dimlessSL}
\end{multline}
where $\kappa_x$ has units of inverse length and $\kappa_E$ has units of inverse energy. In addition, recall that none of the local topological markers can change their value without $\mu_{(\vec{x},E)}^{\textrm{C}} = 0$; as such, it is useful to understand the structure of the locations in $(\vec{x},E)$-space for which the local gap closes $\mu_{(\vec{x},E)}^{\textrm{C}} \rightarrow 0$, see Fig.~\ref{fig:kappa}. If $\kappa_x = 0$, the spectral localizer is block diagonal,
%$\big[\begin{smallmatrix}\kappa_E(H-E\mathbf{1}) & 0 \\ 0 & -\kappa_E(H-E\mathbf{1}) \end{smallmatrix}\big]$
and thus $\textrm{sig}[L_{(\vec{x},E)}^{(\textrm{2D})}(X,Y,H; 0, \kappa_E)] = 0$ for any choice of $(\vec{x},E)$. In this case, the spectrum of the spectral localizer will be determined entirely by the Hamiltonian's spectrum shifted by $E$, and those choices of $(\vec{x},E)$ that result in $\mu_{(\vec{x},E)}^{\textrm{C}} = 0$ will form flat planes in position-energy space for $E \in \spec[H]$, see Fig.~\ref{fig:kappa}a. Conversely, if $\kappa_E = 0$, the spectral localizer is off-block diagonal, again resulting in a balanced spectrum regardless of $(\vec{x},E)$, with $\textrm{sig}[L_{(\vec{x},E)}^{(\textrm{2D})}(X,Y,H; \kappa_x, 0)] = 0$. In this second case, as $X$ and $Y$ are diagonal, the locations in $(\vec{x},E)$-space where $\mu_{(\vec{x},E)}^{\textrm{C}} = 0$ form vertical lines intersecting the coordinates of each lattice site and are independent of $E$, see Fig.~\ref{fig:kappa}e. Thus, in either limit, the spectral localizer is always boring, simply returning information about either the Hamiltonian or the position operators and maintaining a balanced spectrum.

In between these two limits, it is reasonable to expect that structures for which the local gap closes in position-energy space interpolate between these two distributions. However, for a system with non-trivial topology, a closed surface in $(\vec{x},E)$-space always appears for which $\mu_{(\vec{x},E)}^{\textrm{C}} = 0$ as part of the interpolative process, see Figs.~\ref{fig:kappa}b-d. For locations outside of the system, or choices of energy outside of the bounded spectrum, the local markers are provably trivial (see Ref.~\citen{loringPseuspectra}, Lemma~7.4). Therefore, if there is a topologically non-trivial region in position and energy inside the system, any path in position-energy space starting in the non-trivial region and ending outside the system must possess at least one location where $\mu_{(\vec{x},E)}^{\textrm{C}} = 0$ so the local markers can change their values, yielding a closed surface.

\begin{figure}[t]
    \centering
    \includegraphics{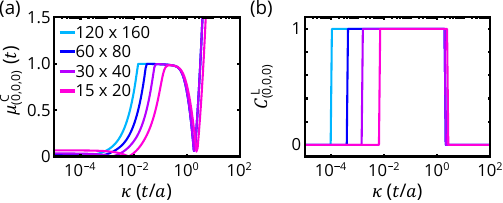}
    \caption{Local gap (a) and local marker (b) for a 2D Haldane model for different $\kappa$ with logarithmic spacing. Calculations use the dimension-full 2D spectral localizer \eqref{eq:2dSL} and are evaluated at the center of the lattice $\vec{x} = \vec{0}$ and the middle of the bulk band gap $E = 0$. Different colors correspond to different lattice sizes, $15 \times 20$ (magenta), $30 \times 40$ (purple), $60 \times 80$ (blue), and $120 \times 160$ (cyan).
    \label{fig:kappa2}}
\end{figure}

For finite systems described by bounded Hamiltonians (i.e., the eigenvalues of finite-sized $H$ are all finite) whose non-trivial topology appears in a bulk spectral gap and stems from the Altland-Zirnbauer classification, there is a proven range over which $\kappa$ is guaranteed to identify the system's non-trivial topology \cite{LoringSchuBa_even}
\begin{align}
    \kappa & \le \frac{E_{\textrm{gap}}^3}{8 \cdot 12 \Vert H_{\textrm{bulk}} - E {\bf 1} \Vert \left(\sum_{j=1}^d\Vert [X_j,H_{\textrm{bulk}}] \Vert \right)}, \\
    \kappa & \ge \frac{E_{\textrm{gap}}}{l}. \label{eq:goodBnd}
\end{align}
Here, $E_{\textrm{gap}}$ is the size of the band gap of the system's bulk Hamiltonian $H_{\textrm{bulk}}$ around the choice of $E$ where the topology is being evaluated, and $l$ the length from the center of the sample to the boundary of the finite system. (Note, relative to the notation used in Ref.~\citen{LoringSchuBa_even} where these bounds are proven, $E_{\textrm{gap}} = 2g$, where $g = \Vert (H_{\textrm{bulk}} - E \mathbf{1})^{-1} \Vert^{-1}$ with $E$ chosen in the center of the bulk spectral gap.) Generally, we find that these bounds are conservative, and even for modestly sized systems choices of $\kappa$ spanning two or more orders of magnitude will produce quantitatively similar results for the spectral localizer's local gap and local markers \cite{dixon_classifying_2023,cerjan_local_2024}. An example comparing the $\kappa$ range correctly classifies a system's topology as the lattice size is increased is shown in Fig.~\ref{fig:kappa2}. Moreover, we typically find that the equality in \eqref{eq:goodBnd} provides an excellent initial choice for $\kappa$, even for systems with unbounded spectra, see also Sec.~\ref{sec:unbounded}.

Finally, as $\mu_{(\vec{x},E)}^{\textrm{C}}$ can change for different choices of $\kappa$, the spectral localizer framework's best estimate for the topological protection at $(\vec{x},E)$ is the maximum local gap over all $\kappa$ that yield the same local markers. Note, different choices of $(\vec{x},E)$ might achieve their largest local gap for different values of $\kappa$. 

\subsection{Overview of multi-operator pseudospectral methods \label{sec:moPseudo}}

In Sec.~\ref{sec:mu}, we introduced the local gap $\mu_{(\vec{x},E)}^{\textrm{C}}$ and discussed how this provided a measure of topological protection for a system. Moreover, in Secs.~\ref{sec:genSL} and \ref{sec:kappa}, we argued that in systems with non-trivial topology the local gap must close everywhere on a closed surface in position-energy space surrounding topological regions in a system. Thus, given that the local gap in a bulk region is quantitatively connected to the bulk spectral gap in that region, it is reasonable to argue that the $(\vec{x},E)$ surface over which $\mu_{(\vec{x},E)}^{\textrm{C}} = 0$ must be near the interface between the topological and trivial regions in a system. Yet, looking at some of the examples considered in this tutorial, such as Figs.~\ref{fig:haldane}c,d, closings of the local gap appear to be co-located with positions of the associated topological edge-localized states in a spectral gap. In this section and the next, we make this connection between a system's LDOS and local gap rigorous, and ultimately show how bulk-boundary correspondence manifests in the spectral localizer framework.

Another way to intuitively understand the mathematical form of the spectral localizer is as the combination of eigenvalue equations, with the form $(M - \lambda)|\phi \rangle = 0$, of the system's Hamiltonian and position operators using a Clifford representation. However, there is an important difference: in standard eigenvalue equations, the eigenvalues are completely determined by the corresponding matrix. Instead, in $L_{(\vec{x},E)}$, $(\vec{x},E)$ can be chosen to be any real numbers and are not limited to those energies in the spectrum of $H$, nor those positions corresponding to lattice vertices. Nevertheless, the resemblance of $L_{(\vec{x},E)}$ to a composition of eigenvalue equations highlights a key feature of the spectral localizer: it can be used to identify approximate joint eigenvectors of the non-commuting constituent matrices, i.e., it can determine if a system exhibits an approximate state $|\phi \rangle$ for which $H |\phi \rangle \approx E |\phi \rangle$ and $X_j |\phi \rangle \approx x_j |\phi \rangle$. Thus, the spectral localizer is an example of a \textit{composite operator}, it could be called the Clifford composite operator, and it defines a \textit{multi-operator $\epsilon$-pseudospectrum}, namely the Clifford $\epsilon$-pseudospectrum. In particular, a system's Clifford $\epsilon$-pseudospectrum is
\begin{equation}
    \Lambda_\epsilon^{\textrm{C}}(\vec{X},H) = \left\{ (\vec{x},E) \left| \; \mu_{(\vec{x},E)}^{\textrm{C}}(\vec{X},H) \le \epsilon \right. \right\}, \label{eq:clifPS}
\end{equation}
i.e., the set of points in position-energy space for which the local gap is less than or equal to $\epsilon$. Note, \eqref{eq:clifPS} is the origin of the superscript in $\mu_{(\vec{x},E)}^{\textrm{C}}$.

Relating a system's Clifford $\epsilon$-pseudospectrum to the appearance of approximate joint eigenvectors is most easily done through the introduction of a second composite operator, the quadratic composite operator \cite{cerjan_quadratic_2022}
\begin{equation}
    Q_{(\vec{x},E)}(\vec{X},H) = \sum_{j=1}^d \kappa^2(X_j - x_j \mathbf{1})^2 + (H - E \mathbf{1})^2.
\end{equation}
Similar to the spectral localizer, the quadratic composite operator defines a local gap,
\begin{equation}
    \mu_{(\vec{x},E)}^{\textrm{Q}}(\vec{X},H) = \left( \min \left[ \spec \left(Q_{(\vec{x},E)}(\vec{X},H) \right) \right] \right)^{1/2}, \label{eq:muQ}
\end{equation}
and the quadratic $\epsilon$-pseudospectrum,
\begin{equation}
    \Lambda_\epsilon^{\textrm{Q}}(\vec{X},H) = \left\{ (\vec{x},E) \left| \; \mu_{(\vec{x},E)}^{\textrm{Q}}(\vec{X},H) \le \epsilon \right. \right\}. \label{eq:quadPS}
\end{equation}
In comparison to the definition of $\mu_{(\vec{x},E)}^{\textrm{C}}$ in \eqref{eq:muC}, the absolute value operation can be dropped in \eqref{eq:muQ} as $Q_{(\vec{x},E)}$ is semi-positive definite, while the square root is needed to adjust $\mu_{(\vec{x},E)}^{\textrm{Q}}$ to have units of energy. Finally, the key reason to introduce the quadratic composite operator is that its local gap is related to the location and localization of an approximate eigenstate across all of the constituent matrices. In particular, it can be proven that (see Ref.~\citen{cerjan_quadratic_2022}, Proposition II.1)
\begin{align}
    \Big(&\mu_{(\vec{x},E)}^{\textrm{Q}} (\vec{X},H) \Big)^2 \notag \\
     =  & \min_{|\phi \rangle \in \mathcal{H}} \bigg\{ \sum_{j=1}^d \kappa^2 \left[ \langle \phi | X_j^2 | \phi \rangle - \langle \phi | X_j | \phi \rangle^2 + \left(\langle \phi| X_j |\phi \rangle - x_j \right)^2 \right] \notag \\
     & \qquad + \langle \phi | H^2 | \phi \rangle - \langle \phi | H | \phi \rangle^2 + \left(\langle \phi| H |\phi \rangle - E \right)^2 \bigg\}, \label{eq:muQbnd}
\end{align}
i.e., the quadratic local gap is equal to the minimum of the bracketed quantity as $|\phi \rangle$ ranges over all of the possible unit vectors in the system's Hilbert space $\mathcal{H}$.
In other words, the quadratic local gap is fundamentally related to whether the system exhibits an approximate joint eigenvector localized near $(\vec{x},E)$ in both its center-of-mass and variances; if $\mu_{(\vec{x},E)}^{\textrm{Q}}$ is small relative to the system's energy scale, such an approximate joint eigenvector exists. Here, there is a critical difference between $\mu_{(\vec{x},E)}^{\textrm{Q}}$ and $\mu_{(\vec{x},E)}^{\textrm{C}}$: the former can only become zero if the constituent matrices at least partially commute and $|\phi \rangle$ is a true joint eigenvector, the latter can become zero even when the constituent matrices do not partially commute. 

More broadly, multi-operator pseudospectral methods are an approach to understanding whether an arbitrary number of non-commuting matrices nevertheless exhibit approximate joint eigenvectors, and a variety of other multi-operator pseudospectra have been proposed \cite{mumford2023numbers_and_beyond,lin2024almost_commuting_and_measurement}. Moreover, the multi-operator pseudospectral methods we discuss here can be considered as a generalization of ``two-operator'' pseudospectra \cite{trefethen_pseudospectra_1997,trefethen_spectra_2005} that have previously been used to study a variety of physical systems \cite{trefethen_hydrodynamic_1993,reddy_pseudospectra_1993,reddy_energy_1993,schmid_linear_2000,baggio_efficient_2020,komis_robustness_2022}. Note though, that both multi-operator pseudospectral methods and two-operator pseudospectral methods are unrelated to ``pseudospectral methods'' as an alternate name for discrete variable representation methods used in the solution of partial differential equations \cite{fornberg_review_1994}.

\subsection{Bulk-boundary correspondence in position space \label{sec:bbc}}

Having discussed how the quadratic local gap is connected to the system's ability to exhibit a localized state, we now turn to the manifestation of bulk-boundary correspondence in the spectral localizer. In particular, notice that the square of the spectral localizer is almost equal to the quadratic composite operator tensored by the identity,
\begin{equation}
    \left(L_{(\vec{x},E)}(\vec{X},H) \right)^2 = Q_{(\vec{x},E)}(\vec{X},H) \otimes \mathbf{1} + \sum_{j=1}^d \kappa [X_j,H] \otimes \Gamma_{j} \Gamma_{d+1},
\end{equation}
where $\Gamma_j$ is the Clifford representation used to define the spectral localizer, and we are using the fact that position operators commute $[X_j,X_l] = 0$. (Alternatively, the spectral localizer can be viewed as almost the square root of the quadratic composite operator using a Clifford representation, much in the same way as how the Dirac equation is related to the Schr\"{o}dinger equation.) As such, the difference between the (Clifford) local gap and the quadratic local gap is bounded,
\begin{equation}
    \left| \left(\mu_{(\vec{x},E)}^{\textrm{C}} (\vec{X},H)\right)^2 - \left(\mu_{(\vec{x},E)}^{\textrm{Q}} (\vec{X},H)\right)^2 \right| \le \sum_{j=1}^d \kappa \Vert [X_j,H] \Vert, \label{eq:bbc}
\end{equation}
see Ref.~\citen{cerjan_quadratic_2022}, Proposition II.4 for a full proof.

Thus, bulk-boundary correspondence naturally appears as a consequence of \eqref{eq:bbc}. As discussed in Sec.~\ref{sec:genSL}, changes in any of the spectral localizer's topological markers can only occur at locations in $(\vec{x},E)$-space where $\mu_{(\vec{x},E)}^{\textrm{C}} = 0$. Moreover, material systems are generally local, such that $\kappa \Vert [X_j,H] \Vert \sim \kappa a$ is small relative to the system's energy scale, where $a$ is the lattice constant. Thus, at locations where a system's local topology changes, \eqref{eq:bbc} guarantees that $\mu_{(\vec{x},E)}^{\textrm{Q}}$ is small, such that there must be a nearby state of the system due to \eqref{eq:muQbnd}. In addition, note that the need for the spectral localizer framework to be applied to finite systems with OBC can be viewed as a consequence of \eqref{eq:bbc}; if PBC were allowed, then $\kappa \Vert [X_j,H] \Vert \sim \kappa l_j$ with $l_j$ the length of the system in the $j$th dimension, which is generally large, and thus a bulk-boundary correspondence would not generally exist.

\begin{figure}[t]
    \centering
    \includegraphics{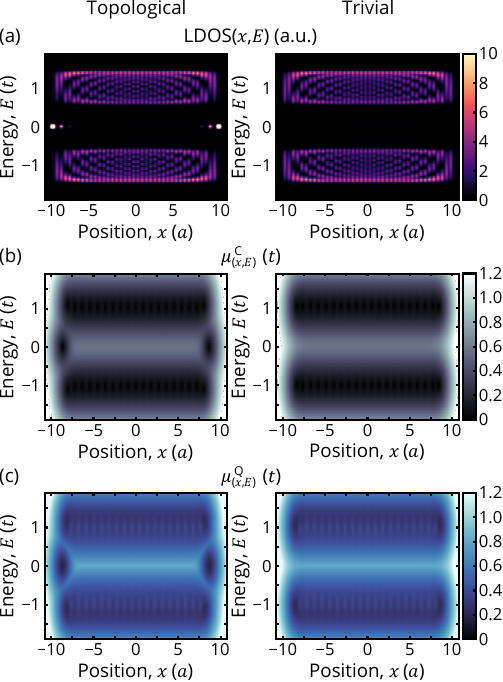}
    \caption{Comparison of the LDOS (a), Clifford local gap (b), and quadratic local gap (c) for a 1D SSH lattice in both topological (left panels) and trivial (right panels) phases with 20 unit cells, coupling coefficients $v = t$ and $w = 0.4t$, and lattice constant $a$. Simulations use $\kappa = 0.5 t/a$.}
    \label{fig:mu}
\end{figure}

However, Eqs.~\ref{eq:muQbnd} and \ref{eq:bbc} have broader utility beyond guaranteeing boundary-localized topological states, they can be used to predict the properties of all of a system's states. For example, the LDOS, local gap, and quadratic local gap are compared for both trivial and topological 1D SSH chains in Fig.~\ref{fig:mu}. As can be seen, both $\mu_{(\vec{x},E)}^{\textrm{C}}$ and $\mu_{(\vec{x},E)}^{\textrm{Q}}$ quantitatively resemble the system's LDOS at all positions and energies. \underline{Open question:} However, note that the (Clifford) local gap becomes smaller than the quadratic local gap where the system exhibits states. Thus, it appears that $\mu_{(\vec{x},E)}^{\textrm{C}}$ should provide a sharper estimate for position-energy locations where a system exhibits an approximately localized state, regardless of whether the state is of topological or trivial origin. Yet, at present, the best known bounds on a state's location and localization are given in terms of $\mu_{(\vec{x},E)}^{\textrm{Q}}$ via \eqref{eq:muQbnd}. Further study is required to understand whether better bounds can be derived in terms of the Clifford local gap.

\subsection{Efficient algorithms and numerical $K$-theory \label{sec:numK}}

It is tempting, but inadvisable, to calculate the spectral localizer's local markers with the forms given in \eqref{eq:markerForms} using a naive approach, e.g., first calculating the determinant or Pfaffian of $L_{(\vec{x},E)}$ and then taking its sign, or first calculating the spectral localizer's full spectrum and then calculating its signature. Unfortunately, even for modestly sized systems, such a naive approach will yield, at best, a slow numerical implementation, and might, at worst, mis-classify the system. Fundamentally, these kinds of approaches fail to take advantage of two separate properties of the spectral localizer framework. First, such naive approaches do not leverage the fact that $L_{(\vec{x},E)}$ is guaranteed to be sparse, which is due to $X_j$ being diagonal and $H$ being reasonably local \cite{kitaev2009}. Second, in viewing the local marker formulae as specifying a sequential algorithm, e.g., find the determinant then take its sign, they miss significant speed-ups available from using matrix factorizations that may not preserve $L_{(\vec{x},E)}$'s spectrum, but do preserve the local marker.

%All the formulas involving the spectral localizer, at least those found so far, have as the final calculation one of three matrix calculations. The matrix involved, basically some part of the spectral localizer with a change of basis or two, is often very large, but also very sparse. Even if the matrix is small, where dense is a good as sparse, one needs to use matrix factorization tricks to avoid a terribly slow algorithm. This is most easily understood in the case of the sign of a determinant. 

An example of the numerical efficiencies available to the spectral localizer framework can be seen in the calculation of the sign of a determinant.
For the sake of specificity, in class D we can change basis so that
the particle-hole operator is just complex conjugation $\mathcal{K}$.
Thus in 1D we have two Hermitian matrices $X$ and $H$ that commute
or anticommute with $\mathcal{K}$. In matrix terms, we have $X$
real (all real matrix entries) and $H$ purely imaginary. In most
models, $X$ will be diagonal, so very sparse, and $H$ will be sparse
or approximable by sparse. This means that if we use
the reduced spectral localizer from \eqref{eq:redSL} with $\mathcal{K}$ instead of $\Pi$,
%\begin{equation*}
%\tilde{L}_{(x,0)}(X,H)=(X-x\mathbf{1})-iH
%\end{equation*}
then $\tilde{L}_{(x,0)}(X,H)$ is real, symmetric, and sparse. The local
index in this case is
\begin{equation}
\textrm{sign}\left(\det\left[ \left(\kappa(X-x\mathbf{1}) - iH\right) \mathcal{K} \right] \right).
\end{equation}

Given a real invertible matrix $M$, we need to take care how we compute
$\textrm{sign}(\det[M])$. Since $\det[M]$ is the product of all the eigenvalues
of $M$, the numerical calculation of this determinant will frequently lead to underflow or overflow errors. Thus, we need to directly compute
the sign of the determinant. Computing all the eigenvalues, taking
sign of each one and taking the product of these $\pm 1$ will avoid the underflow and
overflow issues. However, this will be too slow for all but small
models. A good algorithm can instead be constructed using the LU decomposition of $M$. (Note, LU is not an acronym, it instead references the formula for this decomposition.)
Standard algorithms, applied to $M$, will return a lower-triangular
matrix $L$ and an upper-triangular matrix $U$ so that $M=LU$. Since
\begin{equation}
\textrm{sign}\left(\det[LU]\right)=\textrm{sign}\left(\det[L]\right)\textrm{sign}\left(\det[U]\right)
\end{equation}
we need only take the sign of all diagonal elements of both $L$ and
$U$ and form the product. If $M$ is sparse, it is best to use a
more sophisticated form \cite{davis2004LU_algorithm} of LU factorization
that factors into more matrices, each matrix in some form that makes
the sign of its determinant quick to compute, where all the factors
are about as sparse as $M$. See the \textsc{Matlab} code in a
repository associated to Loring
 \cite{loringPseuspectra,loring_k-theory_2015} for more details. 

We can also compute signatures and signs of Pfaffians efficiently, without total knowledge of the spectrum of $L_{(\vec{x},E)}$ by using similar matrix factorizations.
In the cases where the index is computed via the signature, one can use the so-called LDLT matrix factorization technique to calculate the signature. (Again, LDLT is not an acronym.) If we have $M=L D L^\dagger$ with $L$ being lower-triangular and $D$ being diagonal, we can use Sylvester's law of inertia \cite{sylvester_xix_1852}
\begin{equation}
\textrm{sig}\left[LDL^{\dagger}\right]=\textrm{sig}\left[D\right] \label{eq:sylv}
\end{equation}
and then just count the positive entries on the diagonal of $D$.

In some software packages, the sparse LDLT decomposition is not implemented in the case of complex matrices.  A work-around exists, which takes advantage of the embedding of the complex numbers into the algebra of two-by-two real matrices,
\begin{equation}
a+ib\mapsto\left[\begin{array}{cc}
a & -b\\
b & a
\end{array}\right].
\end{equation}
Thus, the formula that can be utilized is 
\begin{equation}
\textrm{sig}\left(M\right)=\frac{1}{2}\textrm{sig}\left[\begin{array}{cc}
\frac{1}{2}\left(M^{*}+M\right) & -\frac{i}{2}\left(M^{*}-M\right)\\
\frac{i}{2}\left(M^{*}-M\right) & \frac{1}{2}\left(M^{*}+M\right)
\end{array}\right].
\end{equation}
(Lurking mathematicians are reminded that $M^{*}$ here denotes taking complex conjugate only.)

For the sign of the Pfaffian of a real, skew-symmetric matrix $M$, we need to employ a skew-LDLT factorization where software is not readily available.  The basic algorithm
has been developed by Duff \cite{duff2009skew_LDLT}.
For dense matrices, we can utilise the standard Hessenberg factorization, $M=U T U^\top$ that has
$U$ real orthogonal and $T$ being skew-symmetric and tri-diagonal due to being skew-symmetric and in Hessenberg form. The sign of the Pfaffian of $T$ can be quickly computed and we can utilize the formula
\begin{equation}
\textrm{sign}\left(\textrm{Pf}\left[ L T L^\dagger\right]\right) = \textrm{sign}\left(\det\left[ L\right]\right)
\textrm{sign}\left(\textrm{Pf}\left[ T \right]\right) .
\end{equation}

Finally, we note that the local gap can be efficiently computed using sparse matrix methods, which can find the single eigenvalue with the smallest magnitude without determining the full spectrum of the spectral localizer.

Altogether, the local topological markers provided by the spectral localizer framework are examples of numerical $K$-theory, which can broadly be defined as the study and development of formula for $K$-theoretic invariants, such as invariants that classify material topology, that are amenable to efficient numerical calculation. Other examples of numerical $K$-theory outside of the spectral localizer framework have been developed by Prodan \cite{prodan2017computational}, Fulga \cite{fulga2012scattering_theory_TI}, and Quinn \cite{quinn2024approx_interface_top_invariants}.

\section{Applying the Spectral Localizer to Photonic Systems \label{sec:photon}}

Having introduced the spectral localizer framework in general, in this section we now turn to its application to realistic photonic systems. Overall, the process of adapting Maxwell's equations for use with the spectral localizer is relatively straightforward, and can be accomplished with standard discretization techniques, including both finite-difference \cite{cerjan_operator_Maxwell_2022,dixon_classifying_2023} and finite-element \cite{wong_classifying_2024} methods. However, in doing so, two issues arise that are not found in the kinds of simple tight-binding models considered in Secs.~\ref{sec:SLintro} and \ref{sec:4}. First, Maxwell's equations are a set of differential equations with an unbounded spectrum, i.e., these equations exhibit an infinite number of eigenvalues that become infinitely large. Thus, prior to discretization, quantities like $\Vert H \Vert$ are undefined. While discretizing the system yields a finite Hamiltonian matrix with a bounded spectrum, $\Vert H \Vert$ is then determined by the modes at the extremes of $H$'s spectrum that are never in the frequency range of interest because they are not well-described by the discretization (if these frequencies are of interest, a finer discretization is needed). As such, the spectral localizer framework's predictions for topological protection (Sec.~\ref{sec:mu}) and bounds on $\kappa$ (Sec.~\ref{sec:kappa}) need to be adapted. Second, photonic systems commonly feature radiative boundary conditions \cite{taflove_advances_2013}, which are non-Hermitian. Thus, the spectral localizer framework's local markers need to be adapted for non-Hermitian systems featuring line gaps.

In this section, we first remind a reader how Maxwell's equations can be reformulated into an ordinary eigenvalue equation and subsequently discretized to be used in the spectral localizer framework. Then, in Sec.~\ref{sec:genEigval} we discuss how to use a generalized eigenvalue problem in the spectral localizer, which is an approach suited for use with finite-element methods. In Sec.~\ref{sec:nonHerm}, we show how to generalize the spectral localizer and local Chern marker to line-gapped non-Hermitian systems to allow for the inclusion of radiative boundary conditions. In Sec.~\ref{sec:unbounded}, we consider how the spectral localizer framework's local measure of protection needs to be altered for systems with unbounded, or effectively unbounded, spectra. Finally, in Sec.~\ref{sec:nonLin} we comment on using the spectral localizer to address nonlinear systems.

\subsection{Review of Maxwell's equations as an eigenvalue equation}

In linear materials with local responses, the source-free Maxwell's equations for time-harmonic fields with a single frequency component $\omega$ are
\begin{subequations} \label{eq:Maxwells}
\begin{align}
&\nabla \times \vec{E}(\vec{x}) = i \omega \overline{\overline{\mu}}(\vec{x},\omega)\vec{H}(\vec{x}), \label{eq:curlE} \\
&\nabla \times \vec{H}(\vec{x}) = -i \omega \overline{\overline{\varepsilon}}(\vec{x},\omega) \vec{E}(\vec{x}), \label{eq:curlH} \\
&\nabla \cdot \left[\overline{\overline{\varepsilon}}(\vec{x},\omega) \vec{E}(\vec{x}) \right] = 0, \label{eq:divE} \\
&\nabla \cdot \left[\overline{\overline{\mu}}(\vec{x},\omega) \vec{H}(\vec{x}) \right] = 0. \label{eq:divH}
\end{align}
\end{subequations}
Here, $\vec{E}(\vec{x})$ and $\vec{H}(\vec{x})$ are the electric and magnetic fields, and $\overline{\overline{\varepsilon}}(\vec{x},\omega)$ and $\overline{\overline{\mu}}(\vec{x},\omega)$ are the electric permittivity and magnetic permeability material responses, respectively, that are typically spatially varying. Rigorously, for material responses to be dispersive (i.e., frequency dependent), the Kramers--Kronig relations require them to also be lossy (i.e., non-Hermitian) \cite{jackson_classical_1998}. Thus, at a given $\vec{x}$, $\overline{\overline{\varepsilon}}(\vec{x},\omega)$ and $\overline{\overline{\mu}}(\vec{x},\omega)$ are both generally $3 \times 3$ complex matrices. Nevertheless, in many cases it is possible to assume that the absorption lines of a material's response are sufficiently narrow and far away from a frequency range of interest so that these tensors can be approximated as Hermitian. Finally, for $\omega \ne 0$, Eqs.~\ref{eq:divE} and \ref{eq:divH} can be derived from Eqs.~\ref{eq:curlE} and \ref{eq:curlH} due to the vector calculus identity $\nabla \cdot \nabla \times \vec{F}(\vec{x}) = 0$ for any vector field $\vec{F}(\vec{x})$, and thus Eqs.~\ref{eq:curlE} and \ref{eq:curlH} provide a complete description of a photonic system's behavior for non-zero frequencies. Altogether, under these assumptions, Maxwell's equations can be can be rewritten as a generalized eigenvalue problem
\begin{equation}
    W \boldsymbol{\uppsi}(\vec{x}) = \omega M(\vec{x},\omega) \boldsymbol{\uppsi}(\vec{x}), \label{eq:genEig1}
\end{equation}
in which,
\begin{subequations}
\begin{align}
    \boldsymbol{\uppsi}(\vec{x}) & = \left[ \begin{array}{c}
   \vec{H}(\vec{x}) \\
    \vec{E}(\vec{x})
    \end{array} \right], \\
    W & = \left[ \begin{array}{cc}
   0 & -i \nabla \times \\
    i \nabla \times & 0
    \end{array} \right], \\
    M(\vec{x},\omega) & = \left[ \begin{array}{cc}
    \overline{\overline{\mu}}(\vec{x},\omega) & 0 \\
    0 & \overline{\overline{\varepsilon}}(\vec{x},\omega) 
    \end{array} \right].
\end{align}    
\end{subequations}

As a differential operator, \eqref{eq:genEig1} is not yet in a form that can be handled numerically in conjunction with the spectral localizer framework. Instead, the system must be discretized via some algorithm to yield a set of vertices $\vec{x}_l$ where the fields are defined, resulting in finite, bounded matrices $W_{\textrm{disc}}$ and $M_{\textrm{disc}}(\omega)$. For example, choosing a basis in which the fields at each coordinate are clustered together,
\begin{equation}
    \boldsymbol{\uppsi}_{\textrm{disc}} = \left[ \begin{array}{c}
    \vec{H}(\vec{x}_1) \\
    \vec{E}(\vec{x}_1) \\
    \vec{H}(\vec{x}_2) \\
    \vec{E}(\vec{x}_2) \\
    \vdots
    \end{array} \right]
\end{equation}
will result in a block diagonal material response matrix
\begin{equation}
    M_{\textrm{disc}}(\omega) = \left[ \begin{array}{ccccc}
        \overline{\overline{\mu}}(\vec{x}_1,\omega) & & & & \\
        & \overline{\overline{\varepsilon}}(\vec{x}_1,\omega) & & & \\
        & & \overline{\overline{\mu}}(\vec{x}_2,\omega) & & \\
        & & & \overline{\overline{\varepsilon}}(\vec{x}_2,\omega) & \\
        & & & & \ddots
        \end{array} \right],
\end{equation}
while the form of the discretized derivative matrix $W_{\textrm{disc}}$ will depend on the details of how the derivatives are treated.
In addition, the chosen boundary condition must be incorporated into $W_{\textrm{disc}}$ and/or $M_{\textrm{disc}}(\omega)$. For example, if the discretization is being performed using a finite-difference approach on a Yee grid \cite{yee_numerical_1966}, perfect electric conductor (PEC, a photonic equivalent of OBC) boundaries can be implemented in $W_{\textrm{disc}}$ as Dirichlet boundary conditions on the relevant electric field components, resulting in Hermitian $W_{\textrm{disc}}$. Alternatively, to properly capture the capacity for radiation in some photonic systems, stretched coordinate perfectly matched layers (SC-PML) can be incorporated into $W_{\textrm{disc}}$, making it non-Hermitian, or impedance matched material absorption can be incorporated into $M_{\textrm{disc}}(\omega)$, similarly making it non-Hermitian. Overall, it is beyond the scope of this tutorial to provide a full review of different discretization methods and associated boundary conditions, instead we assume an interested reader will already be familiar with such techniques. Altogether, through a chosen discretization and boundary condition, \eqref{eq:genEig1} is transformed to
\begin{equation}
    W_{\textrm{disc}} \boldsymbol{\uppsi}_{\textrm{disc}} = \omega M_{\textrm{disc}}(\omega) \boldsymbol{\uppsi}_{\textrm{disc}}, \label{eq:genEig2}
\end{equation}
which is almost in a form that can be used with the spectral localizer framework.

Note, here we have reviewed Maxwell's equations from the perspective of coupled first-order differential equations. However, it is also possible to use the second-order differential equation form in the spectral localizer framework \cite{cerjan_local_2024}. Indeed, the second-order form has some advantages for some forms of topology as it does not require handling both ordinary vectors like the electric field $\vec{E}(\vec{x})$ and co-vectors (also called pseudovectors) like the magnetic field $\vec{H}(\vec{x})$ in the same eigenvalue equation as they transform differently under some symmetries.

At this point, there are two different ways to proceed. One can reformulate \eqref{eq:genEig2} as an ordinary eigenvalue equation, which preserves $\omega$ as the eigenvalue but places requirements on the constituent materials. Alternatively, one can work with \eqref{eq:genEig2} directly, which does not have any material constraints, but slightly alters how the frequency eigenvalue is handled.

\subsection{Ordinary eigenvalue problem approach}

To incorporate \eqref{eq:genEig2} into the spectral localizer framework as an ordinary eigenvalue equation, we assume that $M_{\textrm{disc}}(\omega)$ is Hermitian and semi-positive definite, such that it exhibits a unique Hermitian semi-positive definite square root $M_{\textrm{disc}}^{1/2}(\omega)$. Physically, this assumption means that all of the system's constituent materials are standard dielectrics, possibly with an anisotropic or gyro-optic response. Numerically, it also forces any radiative boundary to be incorporated in $W_{\textrm{disc}}$. Moreover, one can prove that $M_{\textrm{disc}}^{1/2}(\omega)$ obeys all of the same unitary and anti-unitary symmetries as $M_{\textrm{disc}}(\omega)$, see Ref.~\citen{cerjan_operator_Maxwell_2022}, Supplemental Material. Thus, using the standard substitution $\boldsymbol{\upphi}_{\textrm{disc}} = M_{\textrm{disc}}^{1/2}(\omega) \boldsymbol{\uppsi}_{\textrm{disc}}$, \eqref{eq:genEig2} can be written as an ordinary eigenvalue equation,
\begin{equation}
    M_{\textrm{disc}}^{-1/2}(\omega) W_{\textrm{disc}} M_{\textrm{disc}}^{-1/2}(\omega) \boldsymbol{\upphi}_{\textrm{disc}} = \omega \boldsymbol{\upphi}_{\textrm{disc}}, \label{eq:ordDisc}
\end{equation}
which can be inserted directly into the spectral localizer framework as $L_{(\vec{x},\omega)}\left(\vec{X},M_{\textrm{disc}}^{-1/2}(\omega) W_{\textrm{disc}} M_{\textrm{disc}}^{-1/2}(\omega)\right)$. Note that in many cases, the frequency dependence of $M_{\textrm{disc}}(\omega)$ can be ignored, either because the materials are not dispersive as the relevant topology is of geometric origin, or because the frequency dependence is approximately constant over the relevant range where the system exhibits non-trivial topology. In practice, we have found the approach of this subsection to be useful in conjunction with finite-difference discretization methods \cite{cerjan_operator_Maxwell_2022,dixon_classifying_2023,cerjan_local_2024}.

Mathematically, there are other possibilities for the properties of $M_{\textrm{disc}}(\omega)$ that would allow for a unique square root to be defined, such as if it is negative semi-definite. Physically, these cases are not especially relevant, as they generally correspond to systems that are completely formed of metals that are likely to be highly absorbing at technologically relevant frequencies.

\subsection{Generalized eigenvalue problem approach \label{sec:genEigval}}

Alternatively, \eqref{eq:genEig2} can be inserted directly into the spectral localizer as a generalized eigenvalue equation. To do so, we first rewrite \eqref{eq:genEig2} as
\begin{equation}
    \left(W_{\textrm{disc}} - \omega M_{\textrm{disc}}(\omega)\right) \boldsymbol{\uppsi}_{\textrm{disc}} \equiv H_{\textrm{eff}}(\omega) \boldsymbol{\uppsi}_{\textrm{disc}} = \alpha \boldsymbol{\uppsi}_{\textrm{disc}}
\end{equation}
where we have introduced $\alpha$ as an eigenvalue of the effective Hamiltonian $H_{\textrm{eff}}(\omega)$. Then, the effective Hamiltonian can be inserted into the spectral localizer, and we simply always choose to probe the system at $\alpha = 0$, i.e., using $L_{(\vec{x},0)}\left(\vec{X},H_{\textrm{eff}}(\omega)\right)$ with the system's frequency dependence always incorporated directly into the effective Hamiltonian. In practice, we have found this generalized eigenvalue problem approach to be useful in conjunction with finite-element methods, see Ref.~\citen{wong_classifying_2024} for more details. 

Note, this approach has some subtleties in directly comparing perturbations in the effective Hamiltonian against the local gap that requires considering $(M_{\textrm{disc}}(\omega))^{-1} W_{\textrm{disc}}$. \underline{Open question:} It may be possible to instead define an alternative local gap based on $\sum_j (X_j - x_j \mathbf{1})M \otimes \Gamma_j + (W_{\textrm{disc}} - \omega M_{\textrm{disc}}(\omega)) \otimes \Gamma_{d+1}$, which is guaranteed to have the same signature as a spectral localizer based on \eqref{eq:ordDisc} and would not involve calculating a matrix inverse.

\subsection{Accounting for non-Hermitian phenomena \label{sec:nonHerm}}

Even in the absence of material absorption, many photonic systems exhibit radiative losses resulting in a Hamiltonian that is non-Hermitian. For example, a topological heterostructure constructed in a photonic crystal slab will exhibit out-of-plane radiation for states whose in-plane momenta $\vec{k}_\parallel$ are above the light line $\omega = c |\vec{k}_\parallel|$ \cite{joannopoulos_photonic_2008}. Mathematically, this creates some challenges for the spectral localizer framework because the kinds of homotopy arguments that the framework relies on typically demand Hermiticity. To illustrate this point, recall that Sec.~\ref{sec:2dSLintro} and Fig.~\ref{fig:sigThm} discussed how two invertible Hermitian matrices can only be connected by a path of invertible Hermitian matrices if they have the same signature. However, just as Fig.~\ref{fig:sigThm}b showed how invertible Hermitian matrices with different signatures can be connected by a path in which some of the Hermitian matrices were non-invertible, a similar connection can be made through a path that contains invertible non-Hermitian matrices. Thus, to handle non-Hermitian physical systems, the underlying homotopy arguments must either be replaced or expanded for the spectra localizer framework to remain applicable.

For the specific case of classifying Chern topology (2D Class A) in non-Hermitian systems, the necessary extension of the spectral localizer framework has been rigorously proven, see Ref.~\citen{cerjan_spectral_2023}. To consider such systems, a few minor alternations are made to the framework. First, the form of the spectral localizer is slightly altered, 
\begin{multline}
    {L}_{(x,y,E)}^{(\textrm{2D, NH})} (X,Y,H) = \\ 
        \left( \begin{array}{cc}
        H - E I & \kappa(X-xI) - i\kappa(Y-yI) \\
        \kappa(X-xI) + i\kappa(Y-yI) & -(H - E I)^\dagger
        \end{array} \right). \label{eq:2dSLnh}
\end{multline}
Second, the matrix signature in the definition of the local Chern marker is calculated using only the real parts of the eigenvalues of ${L}_{(x,y,E)}^{(\textrm{2D, NH})}$, i.e., as the difference in the number of eigenvalues with positive real parts and negative real parts,
\begin{equation}
    C_{(x,y,E)}^{\textrm{L}}(X,Y,H) = \frac{1}{2}\textrm{sig}_{\mathbb{R}} \left[ L_{(x,y,E)}^{(\textrm{2D, NH})}(X,Y,H)\right]. \label{eq:SLchernNH}
\end{equation}
Note that this altered index formula is not equivalent to calculating \eqref{eq:SLchern} with the Hermitian portion of \eqref{eq:2dSLnh}.
Finally, the local gap is also only dependent on the real parts of eigenvalues of the non-Hermitian spectral localizer,
\begin{equation}
    \mu_{(x,y,E)}^{\textrm{C}}(X,Y,H) = \min \left( \left| \textrm{Re} \left\{ \spec \left[L_{(x,y,E)}^{(\textrm{2D, NH})}(X,Y,H) \right] \right\} \right| \right).
\end{equation}
As can be seen, all three of these formula reduce to their standard forms from Sec.~\ref{sec:SLintro} as Hermiticity is restored. Intuitively, these formula are taking advantage of the fact that the non-Hermitian system, and its resulting non-Hermitian spectral localizer, remain line gapped (as opposed to point gapped) for Chern topology \cite{gong_topological_2018,kawabata_symmetry_2019,bergholtz_exceptional_2021,okuma_non-hermitian_2023,yang_homotopy_2024} and thus the associated local topological markers can leverage this spectral gap.

Numerically, the switch to non-Hermitian systems also yields problems, as $\textrm{sig}_{\mathbb{R}}[M]$ for non-Hermitian $M$ can no longer be calculated using the LDLT decomposition and Sylvester's law of inertia as detailed in Sec.~\ref{sec:numK} and \eqref{eq:sylv}. Instead, progress has been made by starting with a Hermitian variant of the system and gradually turning on the absorbing boundary condition while monitoring $\mu_{(x,y,E)}^{\textrm{C}}$, which is still efficient to calculate using sparse methods. In the Hermitian variant, the topology can be quickly calculated using techniques from Sec.~\ref{sec:numK}, and then if the local gap remains open as the absorbing boundary condition is turned on, the topology cannot change. Indeed, this approach has proven sufficient in realistic photonic systems using both finite-difference \cite{dixon_classifying_2023} and finite-element \cite{wong_classifying_2024} discretizations.

\underline{Open questions:} There remains a great deal of work to do in this area. First, it is not known whether the local markers for any other class of topology in the Altland-Zirnbauer symmetry classification can be similarly generalized to non-Hermitian systems. Likewise, the non-Hermitian generalization of crystalline topology is not known. This problem is particularly acute for the odd-dimensional classes of topology, where the local markers for Hermitian systems are dependent on a symmetry reduced spectral localizer such as \eqref{eq:redSL} that only includes a single copy of the system's Hamiltonian. In addition, it is unknown whether there are efficient numerical approaches for directly calculating \eqref{eq:SLchernNH}.

The spectral localizer framework has also been extended to consider other forms of non-Hermitian topology by Fulga and colleagues \cite{liu_mixed_2023,ochkan_non-hermitian_2024,chadha_real-space_2024}.

\subsection{Local protection for unbounded operators \label{sec:unbounded}}

Using the spectral localizer framework on systems described by differential equations with unbounded spectra, i.e., systems that exhibit an infinite number of eigenenergies that become infinitely large, presents a problem for the framework's measure of topological protection. Regardless of which specific approach is taken to insert the system's Hamiltonian into the spectral localizer, the norm of any perturbation diverges. Heuristically, this problem stems from the fact that the differential operator $W$ is unbounded, and any perturbation to the material response matrix $M(\omega) \rightarrow M(\omega) + \delta M (\omega)$ still ends up multiplying $W$ for comparison with the local gap, so the full Hamiltonian perturbation remains unbounded. Thus, the criteria for the system to change its local topology $\Vert \delta H \Vert > \mu_{(\vec{x},E)}^{\textrm{C}}(\vec{X},H)$ is always trivially satisfied. Intuitively, the problem is that $\mu_{(\vec{x},E)}^{\textrm{C}}$ is a provably local quantity in both $\vec{x}$ and $E$ (see Ref.~\citen{loring_guide_2019}, Section 7), reflecting whether the system exhibits a nearby state approximately localized at those position-energy coordinates (as discussed in Sec.~\ref{sec:bbc}). In contrast, $\Vert \delta H \Vert$ is a global quantity, and is controlled by the system's response at high energies.

One partial solution to this issue is to project the perturbation into the subspace formed by the $J$ eigenvectors whose corresponding eigenenergies are closest to $E$ where the local gap is calculated. Then, for $J$ chosen to include the states of the finite system that correspond to the neighboring few bands both above and below $E$, the system has an approximate bound 
\begin{equation}
    \Vert \Psi^\dagger \Delta H \Psi \Vert \gtrsim \mu_{(\vec{x},E)}^{\textrm{C}}(\vec{X},H)
\end{equation}
for the perturbation to change the system's local topology \cite{dixon_classifying_2023}. Here, $\Psi$ is the rectangular matrix formed of these $J$ eigenvectors. This approach is also similar to spectral truncation introduced for systems described by differential operators in time \cite{qi_real-space_2024}.

\underline{Open questions:} However, there is a pressing need to develop an exact bound for the topological protection of realistic systems, possibly in terms of a resolvent, rather than such an approximate bound. Indeed, one of the strengths of the spectral localizer framework is its ability to be applied directly to experimentally realizable systems, but those systems are described by wave equations where the difficulty discussed in this section will arise. Ideally, this bound would be formulated in terms of quantities that can be efficiently computed for sparse matrices. Alternatively, the introduction of an alternative local gap, as discussed in Sec.~\ref{sec:genEigval}, may also solve the difficulty discussed in this section by separating out the perturbation to the material response matrix yielding a bound on $\Vert \delta M \Vert$, which is generally finite.

\subsection{Classifying Local Nonlinear Topology \label{sec:nonLin}}

Nonlinear topological systems represent an exciting frontier in photonics, providing a path to exploring phenomena beyond what can be found in electronic topological materials \cite{smirnova_nonlinear_2020}, such as bulk \cite{lumer_nonlinearly_2013, marzuola_bulk_2019, mukherjee_observation_2020, jurgensen_quantized_2021, jurgensen_chern_2022, ren_observation_2023, jurgensen_quantized_2023,jorg_optical_2024} and edge \cite{leykam_edge_2016, smirnova_topological_2019, mukherjee_observation_2021} solitons with topological properties, nonlinearly induced topological phase transitions \cite{hadad_self-induced_2016, zhou_optical_2017, chaunsali_self-induced_2019}, and topological multi-wave mixing \cite{pilozzi_topological_2017, zhang_coupling_2019, mittal_topological_2021, jia_disordered_2023,flower_observation_2024}. Moreover, the spectral localizer framework and its local picture of material topology appears to be extremely well positioned to study these systems \cite{wong_probing_2023,bai_arbitrarily_2024}, as its local markers are able to resolve a system's local change in topology due to its occupation. 

Nonlinear photonic systems are typically considered using a Gross-Pitaevskii equation \cite{gross_structure_1961,pitaevskii_vortex_1961} that describes the effects of particle interactions mediated by an ambient material response in the mean-field limit. In steady-state, such systems are characterized by a nonlinear eigenvalue equation,
\begin{equation}
    H(\boldsymbol{\uppsi})\boldsymbol{\uppsi} = E_{\textrm{NL}}\boldsymbol{\uppsi}, \label{eq:nonlin}
%\equiv \left(H_0 + H_{\textrm{NL}} (\boldsymbol{\uppsi}) \right) \boldsymbol{\uppsi} 
\end{equation}
%where we have separated the linear $H_0$ and nonlinear $H_{\textrm{NL}} (\boldsymbol{\uppsi})$ portions of the Hamiltonian. 
As such, no alterations are needed to the spectral localizer framework to handle such mean-field nonlinearities and identify nonlinear topological phenomena; for any specified occupation $\boldsymbol{\uppsi}$, the system's local topology and associated protection can be found using $L_{(\vec{x},E)}(\vec{X},H(\boldsymbol{\uppsi}))$, see Ref.~\citen{wong_probing_2023}. Moreover, the spectral localizer framework enables a rigorous definition of \textit{topological dynamics}, as it can classify changes in a system's local topology as a system's occupation $\boldsymbol{\uppsi}(t)$ changes in time $t$.

While the definition of local topological protection provided by $\mu_{(\vec{x},E)}^{\textrm{C}}$ remains unchanged for nonlinear systems, it acquires new physical meaning for such systems \cite{wong_probing_2023}. As discussed in Sec.~\ref{sec:mu}, for any perturbation to a linear system described by a bounded Hamiltonian $H \rightarrow H + \delta H$ to change the local topology at $(\vec{x},E)$, the perturbation must be at least strong enough to close the local gap $\Vert \delta H \Vert > \mu_{(\vec{x},E)}^{\textrm{C}}(\vec{X},H)$. However, in nonlinear systems, a small perturbation to the Hamiltonian can change whether a given solution to the nonlinear eigenvalue equation exists. In other words, given a self-consistent solution to \eqref{eq:nonlin}, one can try to follow this solution as the strength of a perturbation is increased, but it is possible that for a sufficiently strong perturbation that the solution curve may simply terminate. Nevertheless, if this self-consistent solution is topological, i.e., it induces a local change in the system's topology at some $(\vec{x},E)$, then $\mu_{(\vec{x},E)}^{\textrm{C}}$ guarantees that the solution curve cannot disappear until the perturbation is strong enough to close the local gap, as the self-consistent solution disappearing causes a change in the system's local topology.

\section{$C^*$-Algebraic Background to the Spectral Localizer \label{sec:whyWorks}}

Overall, this tutorial has focused on providing an understanding of how the spectral localizer framework is used to classify material topology and its associated robustness in physical systems, with the goal of providing a reader with the necessary equations and sufficient intuition to analyze their system of interest. Nevertheless, in this penultimate section of the tutorial, we turn now to providing some explanation of the mathematical underpinnings of the spectral localizer framework. This section is primarily intended for a reader with a mathematics background, with the goal of providing some guidance on the relevant concepts needed to advance the mathematics of the spectral localizer. A reader uninterested this topic may safely skip this section.

\subsection{$C^*$-algebras}

Those familiar with band theory might not realize that the momentum-space
picture $\hat{H}(e^{i\mathbf{k}})=H(\mathbf{k})$, of a tight-binding
Hamiltonian $H$, is an element of a $C^{*}$-algebra. Specifically,
$\hat{H}$ is an element of $C(\mathbb{T}^{d},\mathbf{M}_{N}(\mathbb{C}))$,
which is the set of continuous functions from the $d$-torus to the
$N$-by-$N$ complex matrices. The elements of $C(\mathbb{T}^{d},\mathbf{M}_{N}(\mathbb{C}))$
naturally act on the momentum Hilbert space $\mathcal{H}_{\textup{m}}=L^{2}(\mathbb{T}^{d})\otimes\mathbb{C}^{N}$,
so we could consider $C(\mathbb{T}^{d},\mathbf{M}_{N}(\mathbb{C}))$
as a subalgebra of $\mathcal{B}(\mathcal{H}_{\textup{m}})$. 
In general, $\mathcal{B}(\mathcal{H})$ denotes the algebra of all bounded linear operators on $\mathcal{H}$. 
However, if one wants to know the spectrum of the original Hamiltonian $H$,
acting on position space $\mathcal{H}_{\textup{r}}=\ell^{2}(\mathbb{Z}^{d})\otimes\mathbb{C}^{N}$
for example, one can instead compute the spectrum of $\hat{H}$, not
as an operator on $\mathcal{H}_{\textup{m}}$, but as an element of
the $C^{*}$-algebra $C(\mathbb{T}^{d},\mathbf{M}_{N}(\mathbb{C}))$.
The spectrum is then simply the union of the spectra of all the matrices $H(\mathbf{k})$. In band theory, this relation of band structure to spectrum is explained by explicit calculations, essentially using
the Plancherel (Fourier) transformation. However, this can also be explained in $C^{*}$-algebra language, as an application of spectral permanence; see, for example,  Murphy \cite{murphy2014C_star_algebras}, Theorem~2.1.11.

More broadly, the most basic $C^{*}$-algebra encountered in physics, and the one most central to numerical studies, is simply the $n$-by-$n$ matrices $\mathbf{M}_{n}(\mathbb{C})$. This is essentially the same
as (isomorphic to) $\mathcal{B}(\mathcal{H})$ with $\mathcal{H}=\mathbb{C}^{n}$.
In finite dimensions (as in, $n$ is finite), boundedness is automatic. Thus, the model used to arrive at a definition
of a $C^{*}$-algebra is just $\mathcal{B}(\mathcal{H})$ equipped with the linear structure, operator composition (essentially matrix-matrix multiplication), the spectral norm, and the adjoint. The spectral norm $\|M\|$ of a finite matrix can be described as either the largest
singular value of $M$, or the maximum of $\|M\mathbf{v}\|$ subject to the constraint $\|\mathbf{v}\|=1$.

The `$*$' in $C^{*}$-algebra refers to taking an adjoint. The use of $T^\dagger$ by physicists and $T^*$ by mathematicians for the adjoint operation is a situation that requires linguistic diplomacy in mixed company.
The other needed operations are the operations of addition, scalar multiplication and multiplication of two elements in the $C^{*}$-algebra.
If $F,G:\mathbb{T}^{d}\rightarrow\mathbf{M}_{N}(\mathbb{C})$ are
two elements in $C(\mathbb{T}^{d},\mathbf{M}_{N}(\mathbb{C}))$ then all operations are defined pointwise in momentum, so
\begin{subequations}
\begin{align}
(F+G)(e^{i\mathbf{k}}) & =F(e^{i\mathbf{k}})+G(e^{i\mathbf{k}}), \\
\|F\| & =\max_{\mathbf{k}}\|F(e^{i\mathbf{k}})\|, \\
F^{\dagger}(e^{i\mathbf{k}}) & =(F(e^{i\mathbf{k}}))^{\dagger}, 
\end{align}
\end{subequations}
for example. There are axioms, more than a dozen, that ensure that calculations in a $C^{*}$-algebra are very similar to calculations on bounded operators that does not involve vectors.

Another instance where a physicist will have implicitly done calculations within a  $C^{*}$-algebra is when functions are applied to an operator. If $H$ is Hermitian
with spectrum $\Omega$, then the $C^{*}$-algebra $C^*(H,\mathbf{1})$, formed by taking limits of expressions like $2H+iH^{3}$, will be isomorphic to $C(\Omega)$,
the algebra of all complex-valued continuous functions on $\Omega$. The isomorphism sets up an intuitive correspondence, sending $H$
to the function $h(\lambda)=\lambda$, and the unitary propagator $e^{itH}$ to the function $u(\lambda)=e^{it\lambda}$. 

An example that was the inspiration behind defining the spectral localizer
is $C(S^{d})$, the $C^{*}$-algebra of all complex-valued continuous
functions on the $d$-sphere. If we regard $S^{d}$ as the unit sphere
in $\mathbb{R}^{d}$, we get $d+1$ coordinate functions $\hat{x}_{j}:S^{d}\rightarrow\mathbb{C}$, defined simply by
\begin{equation}    
\hat{x}_{j}(x_{1},\dots,x_{d+1})=x_{j}.
\end{equation}
One then finds that $L_{\boldsymbol{0}}(\hat{x}_{1},\dots,\hat{x}_{d+1})$
is the matrix that determines a generator of $K_{j}(C(S^{d}))$, the $j$th $K$-theory group of the commutative $C^*$-algebra $C(X)$. (One
might need to apply a shift and scaling to this if using (generic) projectors $p^{2}=p^{\dagger}=p$ to build $K_{0}$ groups, as one does classically.)
This is not so obvious for $d\geq4$, but Schulz-Baldes has proven
all the needed details \cite{schulz2023generators_K-groups}.

The most common mapping from one $C^{*}$-algebra to another, say $\varphi: \mathcal{A}\rightarrow\mathcal{B}$, is a $*$-homomorphism. The required conditions on $\varphi$, so it can call itself a $*$-homomorphism, are those that make it linear, along with $\varphi(AB)=\varphi(A)\varphi(B)$,
and $\varphi(A^{\dagger})=\varphi(A)^{\dagger}$. A simple calculation shows that the $*$-homomorphisms $C(S^{2})\rightarrow\mathbf{M}_{n}(\mathbb{C})$ 
can be easily classified. Indeed, if we let $X=\varphi(\hat{x})$,
$Y=\varphi(\hat{y})$, $Z=\varphi(\hat{z})$, then we have the relations
\begin{equation}
\begin{gathered}X^{\dagger}=X,\ Y^{\dagger}=Y,\ Z^{\dagger}=Z,\\{}
[X,Y]=0,\ [X,Z]=0,\ [Y,Z]=0,\\
X^{2}+Y^{2}+Z^{2}=\mathbf{1}.
\end{gathered}
\label{eq:Relations_for_sphere}
\end{equation}
 Any two such triples of matrices can be connected by a path of such
triples, and so we are not surprised to find that $L_{\boldsymbol{0}}(X,Y,Z)$
has always signature equal to zero here. 

Different sorts of mappings $\varphi: \mathcal{A}\rightarrow\mathcal{B}$ started generating interest in the late 1980s \cite{connes_higons1990deform_asymtotiques,lin1999almost_multiplicative}, where one loosens the strict algebraic requirements to allow something like
\begin{equation}
\|\varphi(A)\varphi(B)-\varphi(AB)\|\leq\epsilon.
\end{equation}
Then one can look at generators of a ``fuzzy sphere'' \cite{Madore1992Fuzzy_Sphere} where we keep the Hermiticity conditions in Eqs.~\ref{eq:Relations_for_sphere}, but relax the others to $\|X^{2}+Y^{2}+Z^{2}-\mathbf{1}\|\leq\epsilon$ and $\|[X,Y]\|\leq\epsilon$, etc; this relaxation leads to situations where $L_{\boldsymbol{0}}(X,Y,Z)$ can have nonzero signature.  Moreover, the relaxation of some of the conditions in Eqs.~\ref{eq:Relations_for_sphere} established the connection between almost commuting matrices and $K$-theory \cite{choi_almost_1988,loring1988K-theory_asymptotic_commuting_matrices}. Some abstract, but simple, mathematics can be used to extend the mapping on $\hat{x}$, $\hat{y}$, and $\hat{z}$ to get a function
\begin{equation}
\varphi:C(S^{2})\rightarrow\mathbf{M}_{n}(\mathbb{C})
\end{equation}
 that is ``almost multiplicative'' and so forth.

Now consider a 2D system, with matrix observables $X$, $Y$ and $H$.
If we let $C_{\min}= \min \left[| \spec \left(L_{\boldsymbol{0}}(X,Y,H)\right) | \right]$
and $C_{\max}=\left\Vert L_{\boldsymbol{0}}(X,Y,H)\right\Vert $,
we can define a function 
\begin{equation}
\varphi:C(\Omega)\rightarrow\mathbf{M}_{n}(\mathbb{C})
\end{equation}
that is ``almost a $*$-homomorphism,'' where $\Omega$ is the subset of
$\mathbb{R}^{3}$ of all points $\mathbf{p}$ with $C_{\min}\leq\|\mathbf{p}\|\leq C_{\max}$.
Thus the two $C^{*}$-algebras involved in this theory are extremely
simple, while the map $\varphi$ between them is complicated and
unfamiliar. However, this map $\varphi$ is really only needed to
develop the pseudospectral theory; the spectral localizer and associated
index formulas can be understood without knowing about $\varphi$.

\subsection{$C^*$-algebras with extra symmetries}

Time-reversal in physics is generally described in terms of a Hilbert
space, as it is implemented by an antiunitary operator $\mathcal{T}:\mathcal{H}\rightarrow\mathcal{H}$.
We consider $H$ to have time-reversal symmetry if $H\circ\mathcal{T}=\mathcal{T}\circ H$.
To make this concept compatible with the $C^{*}$-algebra picture,
where the Hilbert space gets shoved to the background, we simply define
a new operation on $\mathcal{B}(\mathcal{H})$
\begin{equation}
A^{\tau}=\mathcal{T}^{-1}\circ A^{\dagger}\circ\mathcal{T}.
\end{equation}
The $\tau$ operation has algebraic properties similar to the transpose operation on matrices. More broadly, the definition of such symmetry-based operations on a $C^{*}$-algebra can be abstracted, resulting in what is called a real
$C^{*}$-algebra, where we have an operation similar to $A\mapsto A^{\tau}$. In momentum space, this operation can look like
\begin{equation}
F^{\tau}(e^{i\mathbf{k}})=F(e^{-i\mathbf{k}})^{\trans},
\end{equation}
where $\trans$ indicates the matrix transpose. There can also be an $N$-by-$N$ unitary $U$ in the definition, specifically 
\begin{equation}
F^{\tau}(e^{i\mathbf{k}})=UF(e^{-i\mathbf{k}})^{\trans}U^{\dagger}.
\end{equation}
This extra operation, and perhaps a second similar operation based on particle-hole conjugation, changes many of the details with defining $K$-theory via vector bundles. Once again,
these challenges are a problem for those developing this theory, but the end result yields explicit formulas that can be understood on their own terms, independent of the theory of real $C^{*}$-algebras.

Consider the system described in Fig.~\ref{fig:ssh_crys}, where only position space is
available. In addition to the Hamiltonian $H$ we have $X$ defined by $X|x_k\rangle=x_k |x_k\rangle$. Both these are elements of the $C^{*}$-algebra $M_{n}(\mathbb{C})$, where $n$ is odd number of sites. Let us here denote the inversion operator by $S$, defined by $S|x_k\rangle= |x_{-k}\rangle$, as discussed in Sec.~\ref{sec:crystal}. We will see that calling inversion $S$ will mathematically conform with the notational conventions of the ten-fold Altland-Zirnbauer classification. We can define a new operation on $M_{n}(\mathbb{C})$ by
\begin{equation}
A^{\sigma}=SAS,
\end{equation}
and adding this structure turns $M_{n}(\mathbb{C})$ into a graded $C^{*}$-algebra.
We call $A$ even if $A^{\sigma}=A$ and odd if $A^{\sigma}=-A$.
We find that $X$ is odd since $XS=-SX$ and $H$ is even since $HS=SH$.
We are in class AIII, but with $H$ in the role usually taken by $X$
and vice versa as compared to the standard case where $S$ denotes chiral symmetry. Notice that in most descriptions of the ten-fold way the position operators go unmentioned, but it is physically reasonable to assume that these commute with any of the local symmetries that are present (i.e., those symmetries used in the standard Altland-Zirnbauer classification).

There is more structure here. Let $\mathcal{T}$ denote complex conjugation,
so bosonic time-reversal $\mathcal{T}^2 = \mathbf{1}$. Let $\mathcal{C}$ denote $\mathcal{T}\circ S$ (where $S$ is still inversion).
Physically, $\mathcal{C}$ is definitely not charge conjugation, but mathematically, we will see that $(H,X)$ can be seen as a pair of Hermitian matrices that are in class BDI, again with roles reversed.  We have $X$ commuting with $\mathcal{T}$ and anticommuting with $\mathcal{C}$ (and so anticommuting with $S=\mathcal{C}\circ\mathcal{T}=\mathcal{T}\circ\mathcal{C}$).
We can use the 1D class BDI index formula \cite{loringPseuspectra}, with the roles of $H$
and $X$ swapped, 
\begin{equation}
\frac{1}{2}\textrm{sig}\left[\left(H-E\mathbf{1}\right)S+\kappa X\right]
\end{equation}
that leads to the same index as the class AIII formula, with the  advantage that $\left(H-E\mathbf{1}\right)S+ \kappa X$ is a real matrix, see Ref.~\citen{cheng_revealing_2023}. In $C^{*}$-algebra terms, we would tend to keep the $A\mapsto A^{\sigma}$ operation and add the operation $A\mapsto A^{\tau}$ where $A^{\tau}=\mathcal{T}\circ A^\dagger\circ\mathcal{T}$.
With the addition of both operations, we have made $M_{n}(\mathbb{C})$ into a graded real $C^{*}$-algebra. 

Theoretical work in the $K$-theory of $C^{*}$-algebras almost always works with homotopy classes of elements that have been \textit{spectrally flattened}. 
That is, one looks at homotopy properies of Hermitian elements with spectrum in $\{-1,1\}$ or unitary matrices, whose spectrum must lie in the unit circle.  However, numerically, spectral flattening is slow and typically results in a dense matrix.  One generally cannot apply formulas from pure math papers on $K$-theory, unmodified, and expect fast algorithms.

\subsection{Clifford algebras}

Clifford algebras operate behind these scenes to help categorize all the possible irreducible Clifford representations \cite{cerjan_even_2024}.  Moreover, it can be useful to consider multiple Clifford representations in the same calculation. For example, two very different Clifford representations can be used to prove that various symmetries in $(\vec{x},E)$ lead to a corresponding symmetry in the Clifford pseudospectrum of $(\vec{x},E)$ \cite{cerjan_even_2024}.

Let $d$ denote one less than the number of Clifford matrices needed, since we usually use $d+1$ Clifford matrices when there are $d$ physical dimensions.
The key thing to know when $d$ is odd is that any two Clifford representations of minimal size will be related via conjugation by a single unitary.  When $d$ is even, this is false, as the Clifford representations come in two flavors. For $d=2$ the two flavors can be distinguished by checking which of $\Gamma_1 \Gamma_2 = \pm i \Gamma_3$ holds.  Within one of the flavors, all irreducible representations are related via conjugation by a single unitary.

The full Clifford  $C^*$-algebra can be described as the minimal $C^*$-algebra containing the universal Clifford representation.  When $d=2$ the relevant Clifford algebra, denoted in math as $\textup{Cl}_{3}(\mathbb{C})$, is a Cartesian product of matrix algebras
\begin{equation}
\textup{Cl}_{3}(\mathbb{C}) = \mathbf{M}_2(\mathbb{C}) \oplus \mathbf{M}_2(\mathbb{C})
\end{equation}
and the universal Clifford representation is
\begin{subequations}
\begin{align}
\hat{\Gamma}_1 & =  \left(\sigma_x,-\sigma_x\right), \\ 
\hat{\Gamma}_2 & =  \left(\sigma_y,-\sigma_y\right), \\
\hat{\Gamma}_3 & =  \left(\sigma_z,-\sigma_z\right).
\end{align}
\end{subequations}
Note, these are tuples of matrices, not matrices. The representation theory of $\textup{Cl}_{3}(\mathbb{C})$ is simple to work out.  Every irreducible Clifford representation is found by selecting a $*$-homomorphism $\pi:\textup{Cl}_{3}(\mathbb{C}) \rightarrow \mathbf{M}_2(\mathbb{C})$ and then setting
$\Gamma_j = \pi(\hat{\Gamma}_j)$. In principle, all our calculations can be done in 
\begin{equation}
\mathbf{M}_n(\mathbb{C})\otimes\textup{Cl}_{d+1}(\mathbb{C}) 
\end{equation}
using the universal Clifford representation, but if we want numerical algorithms, we will want to use an irreducible Clifford representation to minimize the computer memory needed.

\section{Summary and Outlook \label{sec:outlook}}

In this tutorial, we have endeavored to provide a physically motivated introduction to the spectral localizer framework to facilitate its use across the community to address challenges at the frontiers of topological photonics. As this framework provides local markers of material topology, and comes equipped with a local measure of protection, it is able to analyze systems that are either difficult or impossible to consider using traditional approaches, such as topological phase transitions induced via local nonlinearities, effects dependent on finite system sizes, and the appearance of topological phenomena despite the absence of a spectral gap. Moreover, due to the mathematical formulation of the framework's invariants and local gap, these quantities can be computed efficiently even for realistic systems governed by differential operators and numerically described using finite-difference or finite-element methods. In addition, framework's generality has also enabled its application in plenty of condensed matter settings \cite{liu_topological_2023,qi_real-space_2024,franca_topological_2024}. %,spataru_efficient_2024
As part of this introduction, we have outlined the mathematical concepts of multi-operator pseudospectral methods, which allow for the prediction of approximate joint eigenvectors of non-commuting matrices and form the basis of bulk-boundary correspondence in the spectral localizer framework; as well as numerical $K$-theory, the concept that underpins the framework's numerical efficiency. Finally, we have provided some guidance to any interested mathematically oriented reader for how to continue to develop the associated possibly real, possibly graded $C^*$-algebras.

Looking forward, substantial opportunities remain in both the development of the spectral localizer framework as well as its application to novel physical systems to predict new phenomena. Throughout this tutorial, we have marked \underline{open questions} where the framework would benefit from additional results proving a generalization to better address significant classes of physical systems. Moreover, the ability to use the spectral localizer framework in conjunction with models of realistic systems beyond photonics, while including the possibility of aperiodicity or disorder, may yield fruitful results across a range of fields of study. In addition, as the field of topological photonics turns to designing novel device architectures where miniaturization is at a premium, the framework may also find utility in providing a better understanding of topological protection in the presence of finite system size effects. However, more broadly, we are hopeful that given a physically motivated introduction to the subject, the community will find even more applications of the spectral localizer framework.

\section*{Acknowledgements}
We would like to thank Wladimir Benalcazar and Stephan Wong for providing feedback on this tutorial. Florian Sterl is credited for the development of Fig.~\ref{fig:scheme}.
%A.C.\ and T.L.\ acknowledge support from the Laboratory Directed Research and Development program at Sandia National Laboratories.
A.C.\ acknowledges support from the U.S.\ Department of Energy, Office of Basic Energy Sciences, Division of Materials Sciences and Engineering.
T.L.\ acknowledges support from the National Science Foundation, grant DMS-2349959. 
%A.C.\ acknowledges support from the U.S.\ Department of Energy, Office of Basic Energy Sciences, Division of Materials Sciences and Engineering. 
This work was performed, in part, at the Center for Integrated Nanotechnologies, an Office of Science User Facility operated for the U.S. Department of Energy (DOE) Office of Science. Sandia National Laboratories is a multimission laboratory managed and operated by National Technology \& Engineering Solutions of Sandia, LLC, a wholly owned subsidiary of Honeywell International, Inc., for the U.S.\ DOE's National Nuclear Security Administration under contract DE-NA-0003525. The views expressed in the article do not necessarily represent the views of the U.S.\ DOE or the United States Government.

\section*{Data Availability}
Data sharing is not applicable to this article as no new data were created or analyzed in this study.

\bibliography{tutorial_bib}

\appendix

\section{Homotopy results \label{app:homotopy}}

Here we discuss the essential results on the homotopy classification
of specific classes of matrices of a fixed size. We start with proving the theorem  in section~\ref{sec:2dSLintro}.
\vspace{5px}
\newline \noindent 
\textbf{Proof of Theorem A}. The argument to show that invertible
Hermitian, $n$-by-$n$ matrices with different signatures cannot be
connected by a Hermitian path that remains invertible is essentially the
argument in the caption to Fig.~\ref{fig:sigThm}. It is possible to find a continuous
path, perhaps by interpolation, of Hermitian matrices between the
two, but at some point along the path an eigenvalue must cross zero
and that leads to non-invertibility.

Let us show that if $H$ is an invertible, Hermitian, $n$-by-$n$ with $p$ positive eigenvalues then there is a continuous path from
$H$ to 
\begin{equation}
K_{p}=\left[\begin{array}{cc}
I_{p} & 0\\
0 & -I_{n-p}
\end{array}\right].
\end{equation}
Any two such matrices can be connected to this one matrix, so we can just travel along one path and then the other in reverse, showing the claimed connectivity. The spectral theorem for Hermitian matrices applies to $H$. This means there is are orthonormal vectors $|\psi_{1}\rangle$
through $|\psi_{n}\rangle$ such that  
\begin{equation}
H|\psi_{j}\rangle=\alpha_{j}|\psi_{j}\rangle.
\end{equation}
By reindexing, we can assume $0<\alpha_{j}$ for $j\leq p$ and $\alpha_{j}<0$ for $j>p$. We can find a path of these scalars over to $1$ for $j\leq p$ and over to $-1$ for $j>p$, without crossing zero, and so define a path of invertible Hermitian matrices from $H$ to $H_{1}$ where
\begin{equation}
H_{1}|\psi_{j}\rangle=\pm|\psi_{j}\rangle
\end{equation}
with the signs starting as $+$ and then being all minus. We can next
find a path of orthonormal bases, from the $|\psi_{j}\rangle$ over to the canonical basis, and so connect $H_{1}$ to $K_{p}$. The connectedness of all orthonormal bases is equivalent to the classical fact that $U(N)$ is a connected group.

Now we get a subtle situation, that of fermionic parity.  Kitaev\cite{kitaev2009} discusses how this applies to 0D systems in Class D.  See also \cite{grabsch2019pfaffian} for a more detailed discussion of fermionic parity for a coupled pair of quantum dots in a superconducting setting.
\vspace{5px}
\newline \noindent \textbf{Theorem D0:} 
Suppose $n$ is even. Two $n$-by-$n$ invertible
Hermitian skew-symmetric matrices $H$ and $H'$ can be connected by a path of invertible Hermitian skew-symmetric matrices if and only
if $\textrm{sign}\left(\textrm{Pf}\left[iH\right]\right)=\textrm{sign}\left(\textrm{Pf}\left[iH'\right]\right)$.
\vspace{5px}
\newline \noindent 
\textbf{Proof}. 
Suppose $H^\top=-H$ and $H$ is Hermitian and invertible.  Notice this also means $H^* = -H$.
Consider any positive eigenvalue $\alpha$ of $H$. Given $H|\phi\rangle=\alpha|\phi\rangle$ we find
\begin{equation}
H|\phi\rangle^{*}=-H^{*}|\phi\rangle^{*}=-\alpha|\phi\rangle^{*}
\end{equation}
which demonstrates that the spectrum of $H$ is symmetric across zero. This means $\det[H]$ is positive when $n$ is a power of four, and negative when $n=2,6,\dots$. However, $\det(iH)$ will always be positive. Therefore the Pfaffian of $H$, being one of the square roots of $\det(H)$, will be real and non-zero so the sign of the Pfaffian makes sense. Since the Pfaffian varies continuously as $H$ varies, the sign of the Pfaffian cannot change along a path of such matrices. Thus $if \textrm{Pf}\left(iH\right)$ and $\textrm{Pf}\left(iH'\right)$
have opposite signs then $H$ and $H'$ cannot be connected in this space of matrices. 

From any positive eigenvalue $\alpha_{j}$ we obtain a pair of vectors with 
\begin{equation}
H\left|\phi_{j}\right\rangle =\alpha\left|\phi_{j}\right\rangle \text{ and }H\left|\phi_{j}\right\rangle ^{*}=-\alpha\left|\phi_{j}\right\rangle ^{*}
\end{equation}
that must be orthogonal as they live is distinct eigenspaces of $H$. We define real vectors $\left|\psi_{j}\right\rangle $ and $\left|\tilde{\psi}_{j}\right\rangle $ as
\begin{align}
|\psi_{j}\rangle & =\frac{i}{\sqrt{2}}|\phi_{j}\rangle^{*}+\frac{-i}{\sqrt{2}}|\phi_{j}\rangle\\
\bigl|\tilde{\psi}_{j}\bigr\rangle & =\frac{1}{\sqrt{2}}|\phi_{j}\rangle^{*}+\frac{1}{\sqrt{2}}|\phi_{j}\rangle
\end{align}
and so get a real orthonomal basis 
$\bigl|\psi_{1}\bigr\rangle,|\tilde{\psi}_{1}\rangle\dots,\bigl|\psi_{m}\bigr\rangle,\bigl|\tilde{\psi}_{m}\bigr\rangle$
where $n=2m$. 
One calculates
\begin{align}
H|\psi_{j}\rangle & =-\alpha_{j}\frac{i}{\sqrt{2}}|\phi_{j}\rangle^{*}-\alpha_{j}\frac{i}{\sqrt{2}}|\phi_{j}\rangle=-i\alpha_{j}\bigl|\tilde{\psi}_{j}\bigr\rangle\\
H|\tilde{\psi}_{j}\rangle & =-\alpha_{j}\frac{1}{\sqrt{2}}|\phi_{j}\rangle^{*}+\alpha_{j}\frac{1}{\sqrt{2}}|\phi_{j}\rangle=i\alpha_{j}\bigl|\psi_{j}\bigr\rangle
\end{align}
and so in this new basis, $iH$ has matrix representation
\begin{equation}
T=\left[\begin{array}{ccccc}
0 & \alpha_{1}\\
-\alpha_{1} & 0\\
 &  & 0 & \alpha_{2}\\
 &  & -\alpha_{2} & 0\\
 &  &  &  & \ddots
\end{array}\right]
\end{equation}
and that matrix always has positive Pfaffian. However, the basis change can alter the sign of the Pfaffian. If we assemble the 
$\bigl|\psi_{j}\bigr\rangle,|\tilde{\psi}_{j}\rangle$
basis into a real orthogonal matrix $O$ then we have $iH=OTO^{\top}$ and 
\begin{equation}
\textrm{sign}\left(\textrm{Pf}\left[iH\right]\right)=
\textrm{sign}\left(\det[O]\right).
\end{equation}
We can gradually flatten the the spectrum of $H$ and so can assume $\alpha_{j}=1$ for all $j$. Any two real orthogonal matrices of the same determinant (either plus one or minus one) can be connected in the group $O(N)$, completing the proof.
\vspace{5px}
\newline \noindent \textbf{Theorem D1:}
 Two $n$-by-$n$ invertible real matrices $A$ and $A'$ can be connected by a path of invertible real matrices if and only if 
 $\textrm{sign}\left(\det\left[A\right]\right)
 =\textrm{sign}\left(\det\left[A'\right]\right)$.
\vspace{5px}
\newline \noindent 
\textbf{Proof}. 
Suppose  $H$ is real and invertible. The determinent
of a real matrix is always real, and cannot be zero when $H$ is invertble.  Since the determinant is continuous as a function of $H$, we see that two such matrices with determinants of opposite sign cannot be connected in this space of matrices.

To prove that all such matrices of a given sign of determinant are homotopic we first need to reduce to the case where $A$ is real orthogonal.
This we do by utilizing the path $A_\tau=A\left(A^{\dagger}A\right)^{-\tau/2}$.
We see that $A_0=A$ and that $A_1$ is real orthogonal. This construction is continuous in $A$, a fact that can be used to show that two real orthonal matrices that are homotopic in the larger space of inverible real matrices must be homotopic in the smaller space. Real orthogonal
matrices form a group $O(n)$ that has two connected components, $SU(n)$ and the space of real orthogonal matrices of determinant minus one.
One can understand this intuitively due to the nature of the eigenvalues of $A$ when $A$ is real orthongonal. The spectrum of $A$ has three parts. There are eigenvalues at $1$ that do not matter in the sign of the determinant. There are conjugate pairs on the unit circle that also do not matter since their product is positive. It is the eigenvalues at $-1$ that can lead to $\det(A)=-1$. Any two of these can be deformed as a conjugate pair that ends up with both as $+1$. It is the solo eigenvalue at $-1$ that cannot be moved.

\end{document}